\documentclass[journal]{IEEEtran}

\pdfoutput=1

\usepackage{graphicx} % for inclusion of figures
\usepackage[caption=false]{subfig} % for subfigures
\usepackage[nospace,sort,adjust,compress]{cite} % for abbreviated and sorted citings
\usepackage[breaklinks=true]{hyperref}
\usepackage{booktabs} % for professional style tables
\usepackage{tabulary} % for better control of tables
\usepackage[cmex10]{amsmath} % for multi-line eq.s and more
\usepackage{amssymb}

\begin{document}

% use linebreaks \\ within the title to get better formatting as desired
\title{Storms in Mobile Networks}

\author{
Gokce Gorbil, Omer H. Abdelrahman, Mihajlo Pavloski and Erol Gelenbe~\IEEEmembership{Fellow,~IEEE}%
\thanks{G. Gorbil, O.H. Abdelrahman, M. Pavlovski and E. Gelenbe are with the Department of Electrical and Electronic Engineering, Imperial College, London, UK, SW7 2AZ, e-mail: \{g.gorbil, o.abd06, m.pavloski13, e.gelenbe\}@imperial.ac.uk}%
\thanks{Manuscript received September 1, 2014; revised xxx xx, 2014. Accepted for publication xxx xx, 2014.}%
\thanks{\textcopyright~2014 IEEE. Personal use of this material is permitted. However, permission to use this material for any other purposes must be obtained from the IEEE by sending a request to pubs-permissions@ieee.org.}%
}

\markboth{IEEE Transactions on Emerging Topics in Computing,~Vol.~xx, No.~x, Month~2015}{Authors \MakeLowercase{\textit{et al.}}: Storms in Mobile Networks}


\IEEEpubid{0000--0000/00\$00.00~\copyright~2014 IEEE}

\maketitle

%----------------------------------------

\begin{abstract}
Mobile networks are vulnerable to signalling attacks and storms that are caused by traffic patterns that overload the control plane, and differ from  distributed denial of service (DDoS) attacks in the Internet since they directly attack the control plane, and also reserve wireless bandwidth without actually using it. Such attacks can result from malware and mobile botnets, as well as from poorly designed applications, and can cause service outages in 3G and 4G networks which have been experienced by mobile operators. Since the radio resource control (RRC) protocol in 3G and 4G networks is particularly susceptible to such attacks, we analyze their effect with a mathematical model that
helps to predict the congestion that is caused by an attack. A detailed simulation model of a mobile network is used to better understand the temporal dynamics of user behavior and signalling in the network and to show how RRC based signalling attacks and storms cause significant problems in the control plane and the user plane of  the network. Our analysis also serves to identify how storms can be detected, and to propose how system parameters can be chosen to mitigate their effect.
\end{abstract}


%----------------------------------------

\section{Introduction}
\label{sec:intro}

\IEEEPARstart{S}{mart devices} have not gone unnoticed by cyber-criminals, who have started to target mobile platforms~\cite{bib:trendmicroPostPCThreats12,bib:kasperskyMalwareEvolution12,bib:feltSurveyMobileMalware11,bib:chandraMobileMalware12}, and subscribers and mobile network operators (MNOs) face new security challenges~\cite{bib:nemesysGHTCE13}, including the identification and mitigation of \textit{signalling attacks and storms}, which overload the control plane through traffic that causes excessive signalling in the network. The susceptibility of mobile networks to such attacks has been identified ~\cite{bib:enckSMSExploit05,bib:serrorPagingOverload06,bib:leeDetectionDoS3G07,bib:ricciatoDoS3G10}, and they have now become a reality that MNOs have to face regularly due to deliberate, malicious actions either by malware running on the smart devices inside the mobile network, or by Internet hosts outside the core network.

Thus signalling attacks and storms are indeed an emerging cyber-security threat in mobile networks, which are a major component of our cyber infrastructure. As we look at the future, we can expect that UMTS and LTE networks will also support major machine-to-machine communications~\cite{bib:talebMachine3GPP12} where the human being is not in the loop to identify and remediate against an apparent attack. In the first instance, we can expect that UMTS will have to be secured against such attacks and into the future that LTE should be an increasing object of studies to detect and mitigate against signalling storms and attacks~\cite{bib:ksentiniOverloadM2M12,bib:changM2MRA14,bib:fuMobilityM2M14}.

The mobile world witnessed its first botnet in 2012~\cite{bib:kasperskyStatistics12}, through which an attacker can disrupt mobile services by a DDoS-like~\cite{bib:gelenbeLoukasDoS07} attack, overloading the control plane of the mobile network through excessive signalling, rather than the data plane as in traditional DDoS attacks in the Internet. The attacker usually compromises a large number of mobile devices forming a mobile botnet~\cite{bib:mullineriBot10}, which can also be leveraged for other malicious activities in addition to launching signalling attacks. Although in principle some of these attacks can be mitigated by smart routing \cite{Sensible03} inside the core network, such facilities are currently not available.

In order to improve the efficiency of the attack, the attacker can actively probe the network in order to infer the network's parameters~\cite{bib:barbuzziParameter3G08,bib:peralaRrcUmts09,bib:qianRadioAlloc10}, and also identify IP addresses at specific locations within the network~\cite{bib:qianRunHide12}. Indeed, a review of 180 MNOs showed that 51\% of them allow mobile devices to be probed from the Internet, by either assigning them public IP addresses, allowing IP spoofing, or permitting mobile-to-mobile probing within the network~\cite{bib:wangMiddlebox11,bib:qianRunHide12}. Smart mobile devices are also increasingly used in emergency management systems, especially in urban environments~\cite{DBES,bib:gelenbeWuSimEvac12,bib:gorbilCOMPJ12}. Thus they are likely to be targeted in conjunction with other physical or cyber attacks in order to further compromise the safety and confidentiality of civilians and emergency responders~\cite{bib:gorbilANT11,bib:gorbilMSWiM12}.

Since the \textit{radio resource control} (RRC) protocol in UMTS and LTE networks~\cite{bib:rrc3GPP,bib:rrcLTE} is susceptible to signalling attacks, the objective of this paper is to analyse the effect of RRC-based signalling attacks and storms in UMTS networks. While earlier work in this area has focused on signalling behavior from an energy perspective~\cite{Haverinen2007,Yeh2009,Schwartz2013}, we hope to provide a greater understanding of the bottlenecks and vulnerabilities in the radio signalling system of mobile networks in order to pave the way for the detection and mitigation of signalling attacks and storms.

For this purpose, we first present a probability model \cite{Acta1976} of signalling state transitions for a single UMTS user, from which we derive analytical results regarding the user's behaviour when attacked and the impact it has on the network. We also present results from simulation experiments, which enable us to clarify the temporal dynamics of user behavior and signalling and to validate the mathematical model. Then we show how certain specific system parameters such as time-outs can be used to lessen or mitigate the effect of storms and signalling attacks.

\IEEEpubidadjcol

\section{Signalling Storms}

Signalling storms are similar to signalling attacks, but they are mainly caused by poorly designed or misbehaving mobile applications that frequently establish and tear-down data connections in order to transfer small amounts of data. Many mobile applications are designed and developed by software companies who mainly have an ``Internet'' background and thus are not familiar with the control plane of mobile networks. They therefore assume that connectivity is a given and design their applications without taking into account the specifics of mobile networks. A good example is the case of an Android VoIP application popular in Japan, which used frequent keep-alive messages even when the users were idle, causing a signalling overload and a major outage in the mobile network~\cite{bib:docomoStorm12}. In a similar incident, the launch of the free version of the Angry Birds application on Android caused excessive signalling load due to the frequent communications generated by the in-game advertisements~\cite{bib:cornerAngryBirds11}. Such problems have prompted the mobile network industry to promote best practices for developing network-friendly applications~\cite{bib:gsmaSmarterApps12,bib:jiantoApp12}.

Some applications, which may not normally generate excessive signalling, go haywire when an unexpected event occurs, such as loss of connectivity to an Internet server. For example, an important feature of smartphones is the ability to receive ``push notifications'' from cloud services in order to notify the user of an incoming message or VoIP call. This feature is enabled by having the mobile device send periodic keep-alive messages to a cloud server. In normal operation, this keep-alive period is a large value, e.g.~5 minutes. However, if for any reason the cloud service becomes unavailable, then the mobile device will attempt to reconnect more frequently, generating significantly higher signalling load than normal in the process as has recently been reported~\cite{bib:reddingBlackout13}.

Signalling storms could also result from large-scale malware infections which target the user rather than the network, but generate excessive signalling as a by-product of malicious activity. Examples of malware that would cause signalling storms if many users are infected are SMS/email spammers, adware, premium service abusers and botclients. All of these malware generate frequent but small amounts of data, requiring repeated signalling to allocate and deallocate radio channels and other resources, and therefore have a negative impact on the control plane of the network. Unfortunately for the MNOs, such malware are among the top threats currently encountered on smartphones and tablets~\cite{bib:feltSurveyMobileMalware11,bib:trendmicroMobileThreats12,bib:zhouAndroidMalware12}.

Recent incidents have shown that the threat of signalling attacks and storms is very real and that they have the potential to cause major outages in mobile networks. Unlike flash crowds which last for a short time during special occasions and events such as New Year's Eve, signalling attacks and storms are unpredictable and they persist until the underlying problem is identified and resolved by the MNO. Considering their impact on the availability and security of mobile networks, it is evident that MNOs have a strong incentive to safeguard their users from malware and to proactively detect and mitigate signalling attacks and storms in order to protect their infrastructure and services~\cite{bib:nemesysGHTCE13,bib:nemesys2ISCIS13}. 

\section{The Radio Resource Control Protocol}
\label{sec:RRC}

\begin{figure}[tbp]
	\centering
	\includegraphics[width=\linewidth]{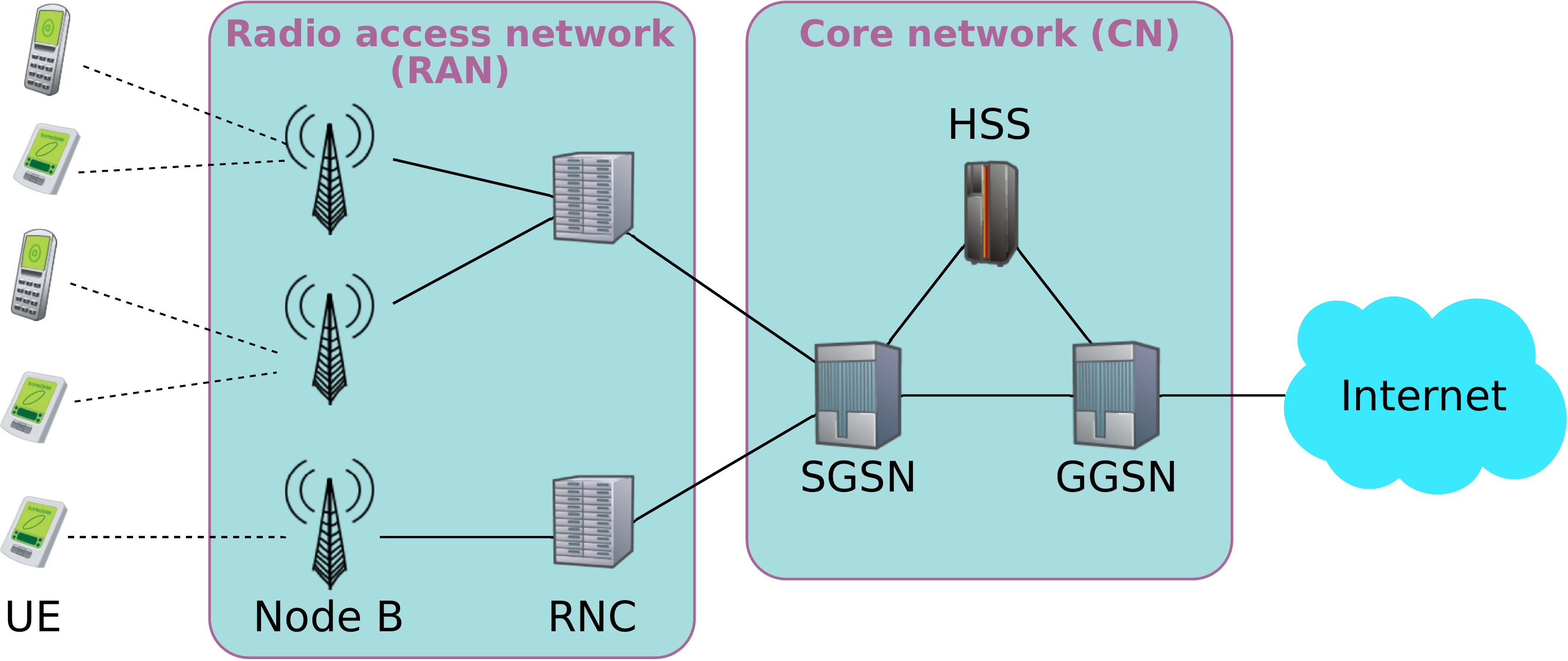}
	\caption{The basic architecture of a UMTS network. UEs are the mobile terminals, e.g.~smartphones, connected to the mobile network via base stations (Node Bs). Node Bs maintain the radio channels with the UEs. The RNC controls the radio resources and the Node Bs in the RAN.}
	\label{fig:umtsNetwork}
\end{figure}

In UMTS networks, the radio resource control (RRC) protocol is used to manage resources in the radio access network (RAN)~\cite{bib:rrc3GPP}. It operates between the UMTS terminals, i.e.~the user equipment (UE), and the radio network controller (RNC). Figure~\ref{fig:umtsNetwork} shows the basic architecture of a UMTS network, depicting the RAN and the core network (CN) elements comprising the packet-switched domain of the mobile network. The RNC is the switching and controlling network element in the RAN. It performs radio resource management (RRM) functions in order to guarantee the stability of the radio path and the QoS of radio connections by efficient sharing and management of radio resources. The RRC protocol is utilized for all RRM-related control functions such as the setup, configuration, maintenance and release of radio bearers between the UE and the RNC. The RRC protocol also carries all non-access stratum signalling between the UE and the CN.

%----------------------------
%----------------------------

\begin{figure}[tbp]
	\centering
	\includegraphics[width=0.65\linewidth]{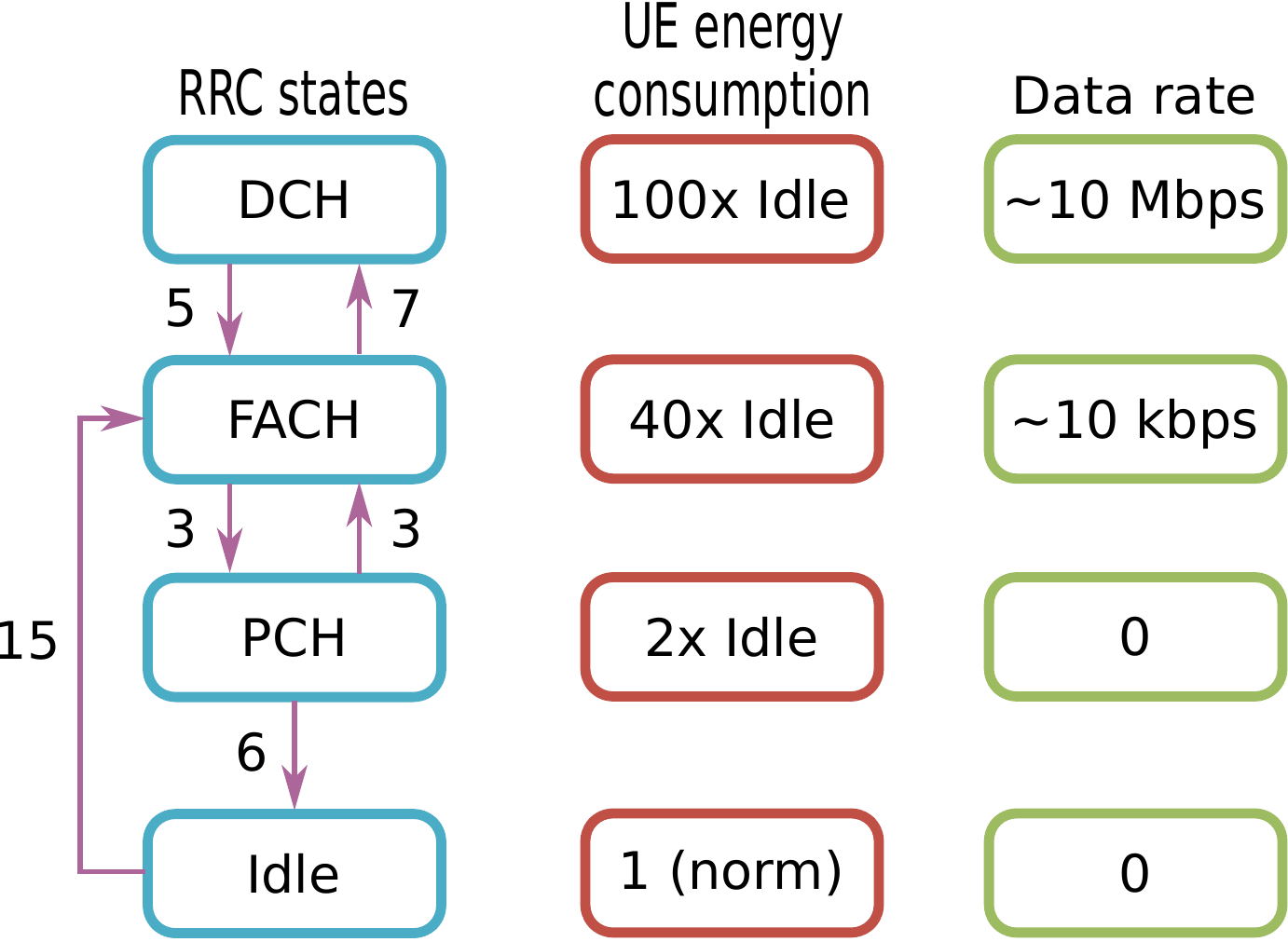}
	\caption{RRC states. The figure on the left shows the typical number of signalling messages exchanged within the RAN for each transition. The other figures show the approximate energy consumption and maximum data rate at the UE.}
	\label{fig:rrcStates}
\end{figure}

In order to manage the radio resources, the RRC protocol associates a \textit{state machine} to each UE, which is maintained synchronized at the UE and the RNC via RRC signalling messages. The RNC controls the transitions between the RRC states based on information it receives from the UEs and the Node Bs on available radio resources, conditions of the currently used radio bearers, and requests for communication activity. As shown in Fig.~\ref{fig:rrcStates}, there are typically four RRC states, given in order of increasing energy consumption and data rate: \textit{idle, cell-PCH, cell-FACH} and \textit{cell-DCH}. In the rest of this paper, we refer to state \textit{cell-X} simply as \textit{X}. Whenever the UE is not in the idle state, it is in \textit{connected mode} and has a signalling connection with the RNC. In connected mode, the location of the UE is known by the RNC at the level of a single cell, which is maintained by \textit{cell updates} sent by the UE either periodically or when it changes cells. We describe the RRC states in more detail below.

%----------------------------
%----------------------------

\textbf{Idle:} This is the initial state when the UE is turned on. In this state, the UE does not have a signalling connection with the RNC, and therefore the RNC does not know the location of the UE. Its location is known by the CN at the accuracy of the location area or routing area, which is based on the latest mobility signalling the UE performed with the CN. Any downlink activity destined for a UE in idle mode will require \textit{paging} in order to locate the UE at the cell level. Since the UE does not have an RNC connection, it cannot send any signalling or data until an RNC connection has been established.

\textbf{FACH:} The UE is in connected mode, and the radio connection between the UE and the RNC uses only common channels which allow low-rate data transmission.

\textbf{DCH:} The UE is in connected mode, and the radio connection uses resources dedicated to the UE. While in DCH, the UE may use shared channels, dedicated channels or both. The data rate of the connection is significantly higher than the FACH state, but energy use is also higher.

\textbf{PCH:} This is a low-energy state that allows the UE to maintain its RNC connection and thus stay in connected mode, but it cannot send or receive any traffic while in this state. While in PCH, the UE listens to paging occasions on the paging channel. This state is optional and it can be enabled or disabled by the MNO according to their policies. Although the PCH state is a low-energy state, the UE still consumes more power than in the idle state. Therefore, some MNOs choose to disable the PCH state in order to allow the UE to return to idle mode quickly and thus reduce its energy consumption. We will investigate the effect of the PCH state on signalling load in Sec.~\ref{sec:results}.

State demotions from a higher to a lower state, e.g.~DCH$\to$FACH, occur based on radio bearer inactivity timers at the RNC. The exact order of state demotions is dependent on MNO policy, but a progression as shown in Fig.~\ref{fig:rrcStates} is common, although some MNOs skip the FACH and/or PCH states. State promotions from the idle and PCH states occur depending on uplink and downlink activity. For example, when the UE has uplink data to send, it sends an ``RNC connection request'' if in idle, or a ``cell update'' if in PCH, to the RNC in order to move to a state where it can send and receive data. Whether the UE is promoted to the FACH or DCH state is dependent on MNO policy. A FACH$\to$DCH transition is performed based on buffer occupancy of the uplink and downlink radio links as observed by the RNC.

Table~\ref{tab:rrcTransitions} summarizes when RRC state transitions occur and the number of signalling messages exchanged to effect each transition. In our simulations, we assume the RRC state progression given in Fig.~\ref{fig:rrcStates}. The UE goes from idle to FACH initially, and then to DCH if the buffer threshold is reached. The UE goes from DCH to FACH upon demotion from DCH. Whether the UE goes from FACH to PCH, or to idle, depends on whether the PCH state is enabled. For an $x \to y$ transition, we use $r_{xy}$ and $c_{xy}$ to denote the number of signalling messages exchanged within the RAN and between the RAN and the CN, respectively.

%----------------------------
% change separation between table columns here; the default is 6pt
\setlength{\tabcolsep}{5pt}

\begin{table}[tbp]
	% IEEE puts table captions before the table:
	\caption{RRC state transitions and number of signalling messages exchanged}
	\label{tab:rrcTransitions}
	\centering
% 	\begin{tabular}{l p{3cm} c c}
	\begin{tabulary}{\linewidth}{LLCC}
		\toprule
		\textbf{Transition} & \textbf{Triggering event} & $\mathbf{r_{xy}}$ & $\mathbf{c_{xy}}$\\
		\midrule
		Idle$\to$FACH	& Uplink or downlink traffic & 15 & 5\\
		PCH$\to$FACH		& Uplink or downlink traffic & 3 & -\\
		FACH$\to$DCH		& Radio link buffer threshold ($\Theta$) reached, $\Theta = 1500$ B & 7 & -\\
		DCH$\to$FACH		& Expiry of inactivity timer $T_1 = 6$s & 5 & -\\
		FACH$\to$Idle	& Expiry of inactivity timer $T_2 = 12$s, PCH disabled & 5 & 3\\
		FACH$\to$PCH		& Expiry of inactivity timer $T_2 = 4$s, PCH enabled & 3 & -\\
		PCH$\to$Idle		& Expiry of inactivity timer $T_3 = 20$min, PCH enabled & 6 & 3\\
		\bottomrule
	\end{tabulary}
% 	\end{tabular}
\end{table}
%----------------------------

The RRC protocol was designed to manage the limited radio resources among multiple UEs and to decrease energy use at the UE. It is therefore biased towards demoting the UE to a lower state as soon as possible, especially if the UE is in the DCH or FACH state. Indeed, as the number of smartphones accessing UMTS networks has increased, the industry has introduced improvements and changes in order to get more data rate out of limited radio resources, such as HSDPA and HSUPA, and to improve the energy use of smartphones. For example, fast dormancy enables the UE to indicate to the RNC when it has no more uplink data to send for a speedier demotion to the PCH or idle state. In addition, some MNOs choose to disable the PCH state in order to allow the UE to return to idle mode quickly and thus reduce its energy consumption. As we will discuss in Sec.~\ref{sec:results}, this tendency to perform hasty RRC demotions result in excessive signalling load in the mobile network, especially in the case of deliberate attacks or signalling storms that result from poorly designed applications.

The RNC will customarily release radio resources for a UE soon after activity ceases in its channel, making those resources available for other UEs. Thus it uses short inactivity timers, which are in the order of 2--10 seconds (see Table~\ref{tab:rrcTransitions}). These short timers make the RRC protocol susceptible to signalling attacks, as an attacker that approximately determines the values of the $T_1$ and $T_2$ timers can then launch a devastating attack from a relatively small number of compromised UEs, as we discuss in Sec.~\ref{sec:results}. In addition, when combined with the ``chatty'' nature of many mobile applications, the tendency to deallocate radio channels quickly necessarily leads to increased RRC signalling in order to reconfigure or setup channels that were released a short time ago, rendering the mobile network vulnerable to RRC based signalling storms.

We thus focus on the RRC protocol in order to better understand its signalling behavior, and investigate under which conditions signalling load becomes excessive. In the next section, we develop a mathematical model of the signalling behavior of the UE, and later derive analytical results from it. Section~\ref{sec:simulation} describes our simulation model of UMTS networks. In Sec.~\ref{sec:results}, we describe our experimental setup and discuss our findings on the effect of signalling attacks targeting the RRC protocol.

\section{Modeling Signalling Behavior of the UE}
\label{sec:mathModel}

Analytical models \cite{Acta1979} are a useful way to gain insight into the main performance interactions within a telecommunications system. Thus we will first review the work in \cite{bib:ICC2014} for a \textit{single} UE's signalling behavior which focuses on the potential of causing signalling storms. We then extend the analysis to include the effect of congestion which limits the signalling load that a set of misbehaving UEs can impose on the network during a storm.

\begin{figure}[tbp]
	\centering
	\includegraphics[width=\linewidth]{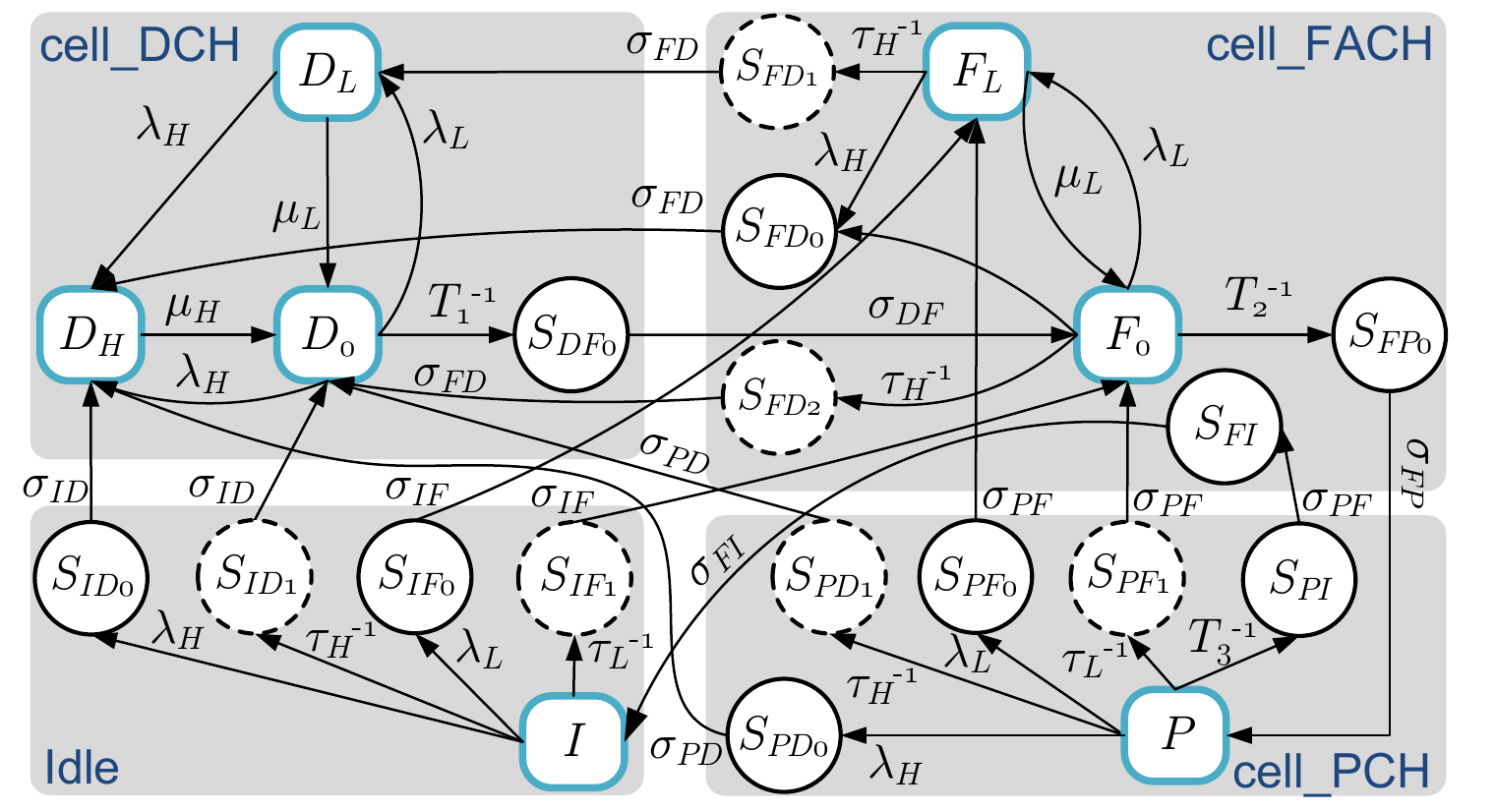}
	\caption{Markov model of the signalling behavior of the UE}
	\label{fig:markovModel}
\end{figure}

Consider a UE which generates both normal and malicious connections, and suppose that its RRC state machine is described by Fig.~\ref{fig:rrcStates}. We will represent the state evolution of the UE by a Markov model as presented in Fig.~\ref{fig:markovModel}. Let $\lambda_L$ and $\lambda_H$ be the rates at which low and high bandwidth connections are {\em normally} made, and $\mu_L$ and $\mu_H$ be the rates at which these connections terminate. Furthermore, denote by $F_L$ the state when the UE is using the bandwidth of FACH, and by $D_L$ and $D_H$ the states when low and high rate requests are handled while the UE is in DCH. Since the amount of traffic exchanged in states $F_L$ and $D_L$ is usually very small, we assume that their durations are independent but stochastically identical. At the end of normal usage, the UE transitions from $F_L$ to $F_0$ or from $D_H,D_L$ to $D_0$, where $F_0$ and $D_0$ are the states when the UE is inactive in FACH and DCH, and before the timers $T_2$ and $T_1$ expire. If the UE does not start a new session for some time, it will be demoted from $D_0$ to $F_0$, and from $F_0$ to $P$, and will then return from $P$ to $I$ (i.e. PCH $\to$ Idle) when inactivity timer $T_3$ expires. Since the UE is not able to communicate in $P$, the transition $P \to I$ is performed by having the UE first move to FACH, release all signalling connections, and finally move to $I$.

%----------------------------
%----------------------------

The attacking or misbehaving connections falsely induce the UE to move from one state to another without the user actually having any usage for such requests. Since in these cases a transition to an actual bandwidth usage state does not take place, unless the user starts a new session, the timers will demote the state of the UE. Consequently, the attack results in the usage of network resources both by the computation and state transitions that occur for session handling, and through bandwidth reservation that remains unutilised.

To perform a signalling attack, the attacker would need to infer the radio network configuration parameters (i.e. the timers $T_i$ and radio link threshold $\Theta$), and also monitor the user's activity in order to estimate when a transition occurs so as to trigger a new one immediately afterwards. Naturally there will be an error between the actual transition time and the estimated one, and we denote the expected value of the difference between the two time instants by $\tau_L$ and $\tau_H$ for malicious transitions to FACH and DCH, respectively. In a similar manner, if the storm is caused by a misbehaving mobile application, then $\tau_L, \tau_H$ represent the level of ``synchronization'' between the malicious traffic bursts and the UE's state changes; for instance $\tau_H = 0$ indicates the extreme case in which a high rate burst is sent immediately after a demotion from DCH.

Finally, let $\sigma_{xy}^{-1}$ be the average time needed to establish and/or release network resources during state promotion or demotion $x\to y$, and $S_{xy}$ be the corresponding state when the UE is waiting in state $x$ for the transition to complete. Note that this overhead is incurred only when the UE moves from one RRC state to another, while changes within the same RRC state (e.g. from inactive to active) occur instantaneously and are seamless to the UE. If $\pi_s$ is the stationary probability that the UE is in state $s$, then the average signalling load (msg/s) on the RNC generated by the UE due to both normal and malicious traffic is:
\begin{align}\nonumber
\gamma_{r}(w) &=~  \pi_I[(\lambda_L+ \tau_L^{-1}) r_{IF} + (\lambda_H+ \tau_H^{-1}) r_{ID}] \\\nonumber
            &+ \pi_P[(\lambda_L+ \tau_L^{-1}) r_{PF} + (\lambda_H+ \tau_H^{-1}) r_{PD}] \\\nonumber
            & + [\pi_{F_0} + \pi_{F_L}] (\lambda_H+ \tau_H^{-1}) r_{FD}   \\\nonumber
            & + \pi_{D_0} T_1^{-1} r_{DF}  + \pi_{F_0} T_2^{-1}[ r_{FP}\mathbf{1_{F\to P}} +  r_{FI} \mathbf{1_{F\to I}} ] \\\label{RAN-rate}
            &+ \pi_{P}T_3^{-1} r_{PI}\mathbf{1_{F\to P}},
\end{align}
where the characteristic function $\mathbf{1_{x\to y}}$ takes the value 1 if the transition $x\to y$ is enabled and 0 otherwise, and $w$ is a congestion parameter which we define in the following section. The UE  also generates signalling with the CN whenever it moves from or to the Idle state, leading to an average signalling load on the SGSN given by:
\begin{align}\nonumber
\gamma_{c}(w) =&~ \pi_{I}[(\lambda_L+ \tau_L^{-1}) c_{IF} + (\lambda_H+ \tau_H^{-1}) c_{ID}] \\\label{CN-rate}
              & +  \pi_{F_0} T_2^{-1} c_{FI} \mathbf{1_{F\to I}} + \pi_{P} T_3^{-1} c_{PI}\mathbf{1_{F\to P}}.
\end{align}

\subsection{Modeling Congestion in the Control Plane}

The analytical model we just described can be solved in closed-form \cite{bib:ICC2014} when the average transition delays are known, allowing to determine the conditions and parameters for which signalling misbehavior has the most serious consequences on the network functioning. In normal circumstances, state promotions and demotions last for few milliseconds that represent only a small fraction of the total lifetime of a session. However, when the mobile network servers become overloaded, as in during a signalling storm, the time needed to establish and release connections also increases, which in turn limits the maximum signalling load that a set of misbehaving UEs can impose on the network. To better understand the effect of a signalling storm, we develop a simple model for the average time $\sigma_{xy}^{-1}$ needed to perform the transition $x \to y$ as follows:
\begin{equation}
\sigma_{xy}^{-1}(w) = r_{xy} w + \sum_{n=1}^{r_{xy}} ~ (t_{xy}[n]  + \delta_{xy}[n])
\end{equation}
which consists of three components:
\begin{itemize}
	\item Communication delay $t_{xy}[n]$ comprising propagation and transmission parts that are subject to the physical characteristics of the links traversed by the $n$-th signalling message exchanged during the transition. This delay depends only on the path followed by the message, and we ignore queueing at the transmission links, since signalling storms do not affect the data plane, and thus they do not translate into congestion in the wireless or wired links.
	
	\item Average queueing delay $w$ at the RNC signalling server, which is a function of the number of normal UEs served by the RNC $M^\mathcal{N}$, the number of misbehaving ones $M^\mathcal{A}$, and the RNC signalling load \eqref{RAN-rate} of both normal $\gamma_{r}^\mathcal{N}$ and misbehaving $\gamma_{r}^\mathcal{A}$ UEs. Note that we do not represent congestion at the SGSN, since the CN is less susceptible to signalling storms, especially when PCH is enabled.
	
	\item Processing time $\delta_{xy}[n]$ at the mobile network servers handling the message, which we assume to be constant per message type\footnote{Note that signalling message types are defined by the 3GPP standards and known a priori.} such that $\delta_{xy}[n]=\sum_{s \in servers}\delta_{xy,s}[n]$.
\end{itemize}

The aggregate load that the RNC signalling server needs to handle is then:
\begin{equation}
\Gamma_{r}(w) = M^\mathcal{N} \gamma_{r}^\mathcal{N} + M^\mathcal{A} \gamma_{r}^\mathcal{A}.
\end{equation}
Note that $\Gamma_r$ is a function of $w$, which itself is determined by $\Gamma_r$. Using a simple $M/M/K$ system to model the RNC signalling server, the average queueing delay can be obtained by solving the non-linear expression \cite{bib:gelenbePujolle98}:
\begin{equation}
w = \frac{(K\rho)^K}{K!(1-\rho)(K\nu - \Gamma_r)} \left[\sum_{i=0}^{K-1} \frac{(K\rho)^i}{i!} + \frac{(K\rho)^K}{K!(1-\rho)} \right]^{-1},
\end{equation}
where $\rho(w) = \frac{\Gamma_r}{K\nu}$, and $\nu$ is an ``equivalent'' average service rate which depends on the composition of the signalling messages processed by the RNC:
\begin{equation}
\nu^{-1}(w) = \Gamma_{r}^{-1}\sum_{\mathcal{C}\in \{\mathcal{N},\mathcal{A}\}}M^{\mathcal{C}}\sum_{x,y} \sum_{n=1}^{r_{xy}} \gamma_{r,xy}^\mathcal{C} \delta_{xy,r}[n],
\end{equation}
where $\gamma_{r,xy}^\mathcal{C}$ is the signalling load on the RNC from a UE of type $\mathcal{C}\in \{\mathcal{N},\mathcal{A}\}$ due to a transition $x\to y$, and $\delta_{xy,r}[n]\geq 0$ is the RNC's processing time of the $n$-th signalling message exchanged during the transition.

\section{Simulation of UMTS Networks and Signalling Anomalies}
\label{sec:simulation}

The mathematical model we have developed and described in Sec.~\ref{sec:mathModel} provides a good approximation of the signalling behavior of the UE, and enables us to quickly derive analytical results in order to investigate the effect of signalling attacks and the values of the various network parameters, such as the $T_i$ timers, on signalling load. In order to capture aspects of the mobile network not explicitly represented in the mathematical model, we have developed a discrete event simulation (DES) model of the UMTS network, focusing on the signalling layer in the RAN. We have developed models of the UE, Node B, RNC, SGSN and GGSN, and also models of the ``Internet cloud'' and Internet hosts (i.e.~servers). While we do not model the circuit-switched (CS) domain explicitly, the SGSN model contains aspects of the MSC server necessary to establish and tear-down CS calls, i.e.~voice calls and SMS; our SGSN model is therefore a hybrid of the SGSN and the MSC server.

The performance of the simulation was an important consideration in our model design, and in order to be able to simulate large scale mobile networks, we have adopted two approaches. First, we have developed our simulation model so that we support \textit{distributed simulation}. We can therefore distribute elements of the simulated mobile network over multiple logical processes in order to leverage multiple hosts in a simulation, allowing us to simulate much larger mobile networks than would be possible with a single process. Second, we combine \textit{packet-level} and \textit{call-level} representation of communications in our model. Communications that are natively message based or bursty in nature are represented at the packet level. These include communications for SMS, email, web browsing, and instant messaging. Other types of communications are represented at the call level; examples include voice calls, VoIP calls, and multimedia streaming.

\begin{figure}[tbp]
	\centering
	\includegraphics[width=0.55\linewidth]{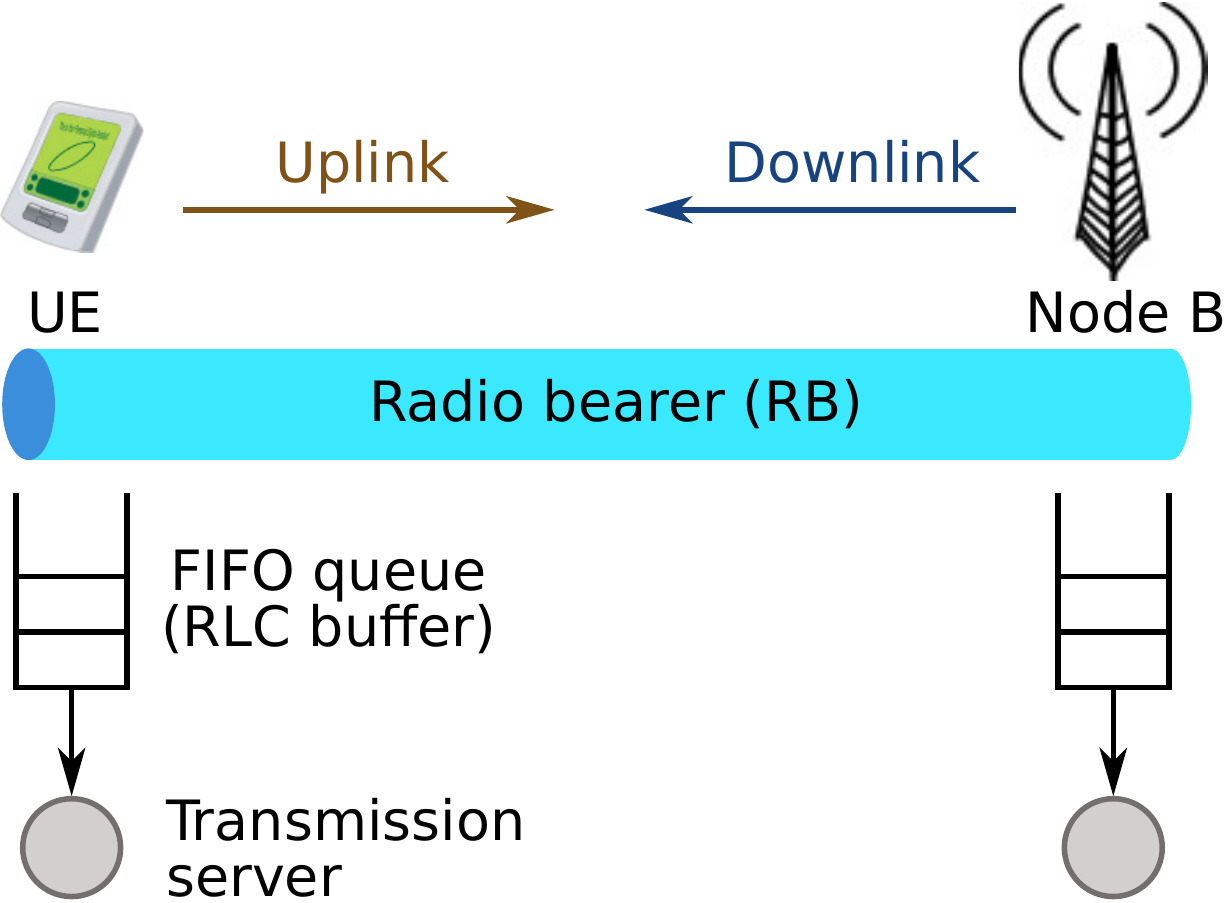}
	\caption{The simulation model of a radio bearer (RB), consisting of a single server, single FIFO queue pair in each direction. The uplink and downlink servers are located at the UE and the Node B, respectively.}
	\label{fig:radioBearer}
\end{figure}

In the control plane, the UE model consists of the session management (SM), GPRS mobility management (GMM) and RRC layers. In the data plane, it contains the application layer, which has CS and IP applications representing all user activity, the transport layer (TCP and UDP) and a simplified IP layer that is adapted for mobile networks. We have a simplified model of the RLC layer, but we do not explicitly model the MAC and PHY layers; effects of changes in radio conditions are modeled as random variations in the data rate of the radio channels. Uplink and downlink radio transmissions over a radio bearer (RB) are modeled by two single server, single FIFO queue pairs, one for each direction as shown in Fig.~\ref{fig:radioBearer}. The service time at the transmission server is calculated based on the length of the currently transmitted RLC packet and the current data rate for the RB. Changes in the RB data rate are reflected on the service time of the current packet. Each UE has one signalling RB and one data RB. In addition to the transmission delays for the RBs, propagation and processing delays are also modeled. We also model the usual communication delays (i.e.~transmission, propagation and processing delays) over wired links connecting the different network elements, e.g.~between the RNC and the SGSN.

%----------------------------
%----------------------------

Our RNC model has the RRC, RANAP, NBAP and GTP protocols. The RRC model in the RNC consists of a single signalling server and a single FIFO queue, used to model the processing time $\delta^{r}_{xy}$ for RRC signalling messages. The server handles two classes of signalling messages, where one class consists of signalling messages that effect a state transition $x \to y$ (e.g.~the RB setup message), and the second class includes all other signalling messages. The service time assigned to the first class reflects the time taken to allocate and deallocate radio resources by the RNC, whereas a default and smaller service time is used for the second class. In the analytical results presented in the next section, $K=1$, and $\nu$ is calculated based on the $\delta^{r}_{xy}$ values as defined here. As the handler of RRC state transitions, this server will be one of the main points of interest in our simulations, and as we discuss in Sec.~\ref{sec:results} it will become overloaded as the severity of the signalling attacks increases.

\section{Experiments}
\label{sec:results}

In order to understand the effect of RRC based signalling attacks in UMTS networks, we implemented our simulation model in the OMNeT++ simulation framework~\cite{bib:vargaOmnetpp08}. We present results from our simulation experiments and analytical results derived from our mathematical model. The UMTS network topology used in the simulations closely resembles the architecture shown in Fig.~\ref{fig:umtsNetwork}. In our simulations we have 1000 UEs in an area of $\mbox{2x2 km}^2$, which is covered by 7 Node Bs connected to a single RNC. The CN consists of the SGSN and the GGSN, which is connected to 10 Internet hosts acting as web servers. All UEs attach to the mobile network at the start of the simulation, and remain attached. We simulate high user activity in a 2.5 hour period, during which users are actively browsing the web. Our web browsing model is based on industry recommendations~\cite{bib:cdmaEval09}, and is described below.

\subsection{Web Browsing Behaviour of the User}

\begin{figure}[tbp]
	\centering
	\includegraphics[width=\linewidth]{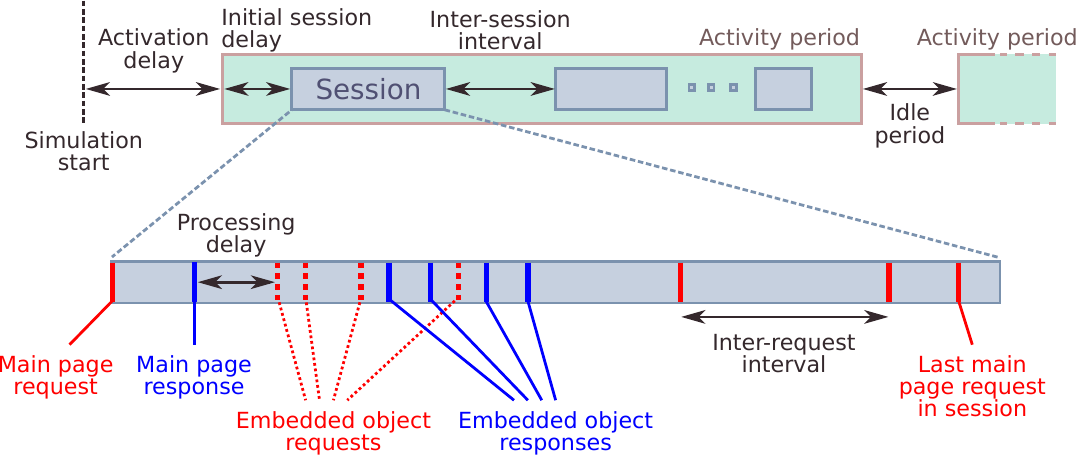}
	\caption{Web traffic model representing interactive user browsing. Note that time is not drawn to scale.}
	\label{fig:webModel}
\end{figure}

We model interactive web browsing behavior using a self-similar traffic model as shown in Fig.~\ref{fig:webModel}. The parameters of the web traffic model are random variables from probability distributions, and Table~\ref{tab:webParams} gives the values used in our simulations, which are based on web metrics released by Google~\cite{bib:googleWebMetrics10}. In addition to this random mode, the web application model also supports a scripted mode in which given user traces are replayed in order to inject browsing events at predetermined times.

%----------------------------
%----------------------------

%----------------------------
% change separation between table columns here; the default is 6pt
\setlength{\tabcolsep}{5pt}

\begin{table}[tbp]
	% IEEE puts table captions before the table:
	\caption{Parameters of the web traffic model}
	\label{tab:webParams}
	\centering
% 	\begin{tabular}{l p{2.87cm} p{4.3cm}}
	\begin{tabulary}{\linewidth}{LLL}
		\toprule
		\textbf{Name} & \textbf{Description} & \textbf{Value}\\
		\midrule
		$p_a$ & Activity period & constant, 24 hours\\
		$d_a$ & Activation delay (min.) & uniform(1, 10)\\
		$d_s$ & Initial session delay & $i_s / 2$\\
		$n_s$ & Number of main page requests in session & truncated normal $\mu=10$, $\sigma=5$, $\mbox{min} = 2$\\
		$i_s$ & Inter-session interval (min.) & truncated normal, $\mu=20$, $\sigma=10$, $\mbox{min} = 2$\\
		$i_r$ & Inter-request interval (sec.) & truncated exponential, $\lambda^{-1} = 60$, $\mbox{min} = 10$, $\mbox{max} = 600$\\
		$l_r$ & Request size (B) & truncated normal, $\mu = 600$, $\sigma = 100$, $\mbox{min} = 300$\\
		$l_m$ & Main page size, excluding embedded resources & histogram~\cite{bib:googleWebMetrics10}\\
		$l_{img}$ & Size of image resources (KB) & truncated exponential, $\lambda^{-1} = 50$, $\mbox{min} = 1.2$, $\mbox{max} = 400$\\
		$l_{txt}$ & Size of text resources & histogram~\cite{bib:googleWebMetrics10}\\
		$n_e$ & Number of embedded objects in page & histogram~\cite{bib:googleWebMetrics10}\\
		$R_{img}$ & Ratio of image resources to all embedded resources in page & uniform(0.1, 0.5)\\
		$d_{p_c}$ & Processing delay, client (ms) & truncated normal, $\mu=50$, $\sigma=10$, $\mbox{min} = 0$\\
		$d_{p_s}$ & Processing delay, server (ms) & truncated normal, $\mu=4$, $\sigma=1$, $\mbox{min} = 1$\\
		\bottomrule
	\end{tabulary}
% 	\end{tabular}
\end{table}
%----------------------------

The activity period represents the time that the UE is active during a 24 hour period, i.e.~the hours during the day that it is generating web traffic. The idle period between two activity periods is the remaining hours within the 24 hours. The first activity period starts after an activation delay $\mathbf{d_a}$, and consists of one or more browsing sessions. The first session within an activity period starts after an initial session delay $\mathbf{d_s}$, and the time between the last and the first main request in one session and the next respectively, is the inter-session interval $\mathbf{i_s}$.

Within a session, the user generates requests for web pages, which are called main page requests, and the first request is scheduled at the start of the session. The request results in a page response from the web server, which is subject to a processing delay $\mathbf{d_{p_c}}$ at the client, representing the time it takes for the web client at the UE to process the received response. A web page contains zero or more embedded objects, and the client generates an embedded object request for each one. We assume that HTTP version 1.1 is used and that each embedded object request is pipelined over a single TCP connection. The length of a request is denoted by $\mathbf{l_r}$. The inter-request interval $\mathbf{i_r}$ is the time between the generation of two consecutive main page requests, and it is independent of the reception of the responses. The session length is controlled by the number of main page requests $\mathbf{n_s}$ in the session.

The web server generates a response for each request it receives after a processing delay $\mathbf{d_{p_s}}$. The length of a main page response is $\mathbf{l_m}$, and it excludes the sizes of any embedded objects and TCP/IP headers. The number of embedded objects per page is $\mathbf{n_e}$ and we model two types of objects: image and text (e.g.~CSS documents, scripts). The size of an embedded object is $\mathbf{l_{img}}$ and $\mathbf{l_{txt}}$ for image and text objects, respectively. $\mathbf{R_{img}}$ gives the ratio of image objects to all embedded objects in a page. In the simulations, a client selects a web server uniformly at random for each main page request.

\subsection{The Attack Model}

We consider two different attack strategies in our evaluation: FACH and DCH attacks. In \textit{FACH attacks}, the attacker aims to overload the control plane by causing superfluous promotions to the FACH state, and therefore needs to know when a demotion from FACH occurs in the UE. In \textit{DCH attacks}, the demotion of interest is from the DCH state. As introduced in Sec.~\ref{sec:mathModel}, the error between the actual transition time and the estimated one is denoted by $\tau_L$ and $\tau_H$ in the FACH and DCH attack scenarios respectively.

In FACH attacks, the attacker sends a small data packet to a random Internet server in order to cause a promotion to FACH. Higher rate data traffic is generated in DCH attacks in order to cause the buffer threshold to be reached and therefore result in a promotion to DCH. For simulation purposes, our RRC model at the UE informs all registered malicious applications when an RRC state transition occurs. Before launching the next attack, the attacker waits for a period of $\tau_L$ or $\tau_H$ after a suitable demotion is detected, e.g.~from FACH to PCH in the FACH attack case, where $\tau_L$, $\tau_H$ are random variables. In our experiments, we assume that $\tau_L$, $\tau_H$ are exponentially distributed with mean = $\{0, 1, 2, 4, 6, 10, 14, 20, 30\} s$ to simulate varying degrees of error on behalf of the attacker. For signalling storms, the $\tau$'s represent the ``synchronization'' between the RRC state machine of the UE and the misbehaving application, while the attack scenario represents whether the misbehaving application generates low-rate or high-rate traffic. We present results from the DCH attack scenario only since the FACH attack scenario produces similar behaviour in most cases.

\section{Modeling and Simulation Results}

\begin{figure}[tbp]
	\centering
	\subfloat[Application response time vs. $1/\tau_H$, for different number of attackers] {
		\includegraphics[width=0.45\linewidth]{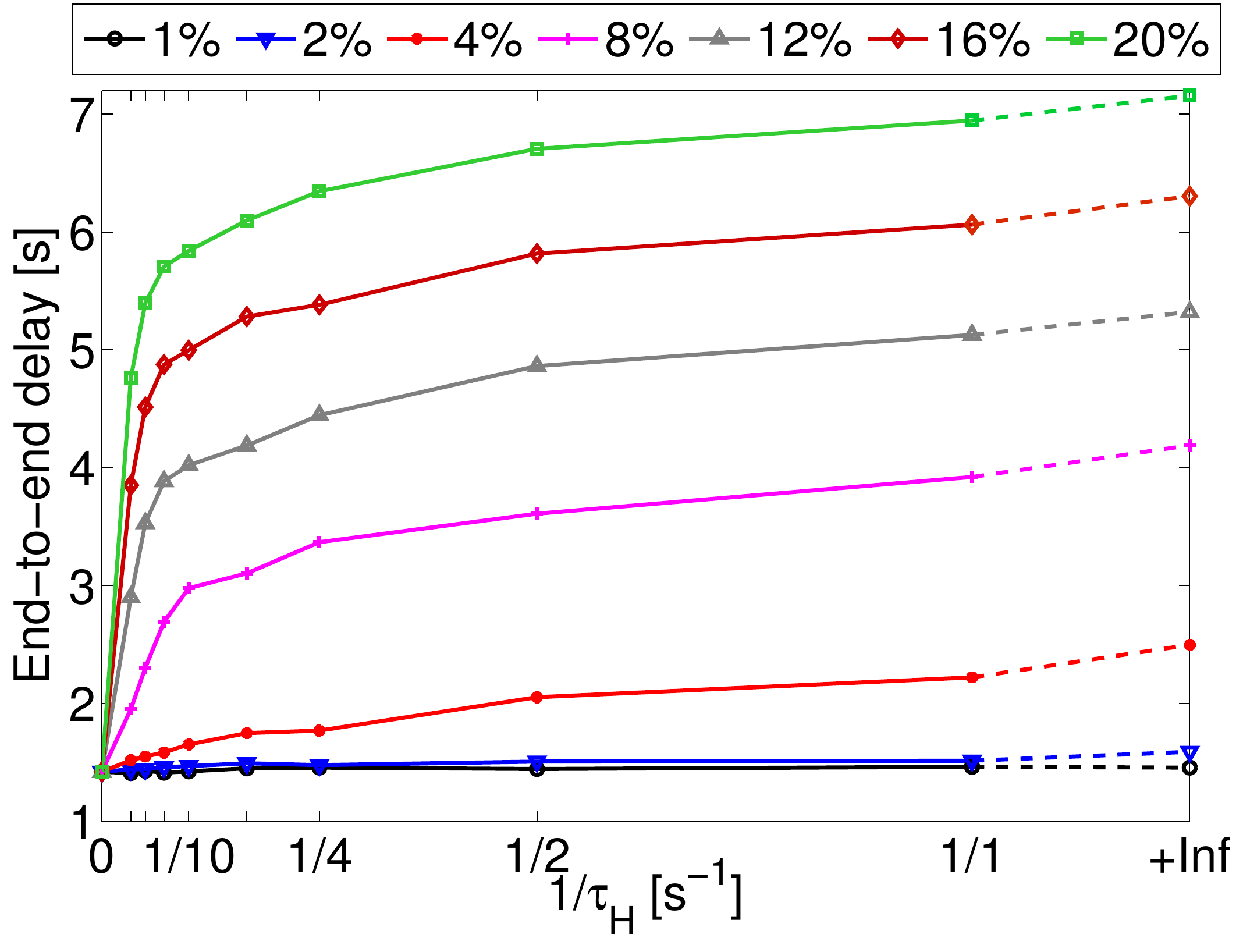}
		\label{fig:dchNoPchResponseTime}
	} \hspace{0.2cm}
	\subfloat[Average queueing time at the RNC signalling server vs. number of attackers, for different $\tau_H$ values] {
		\includegraphics[width=0.45\linewidth]{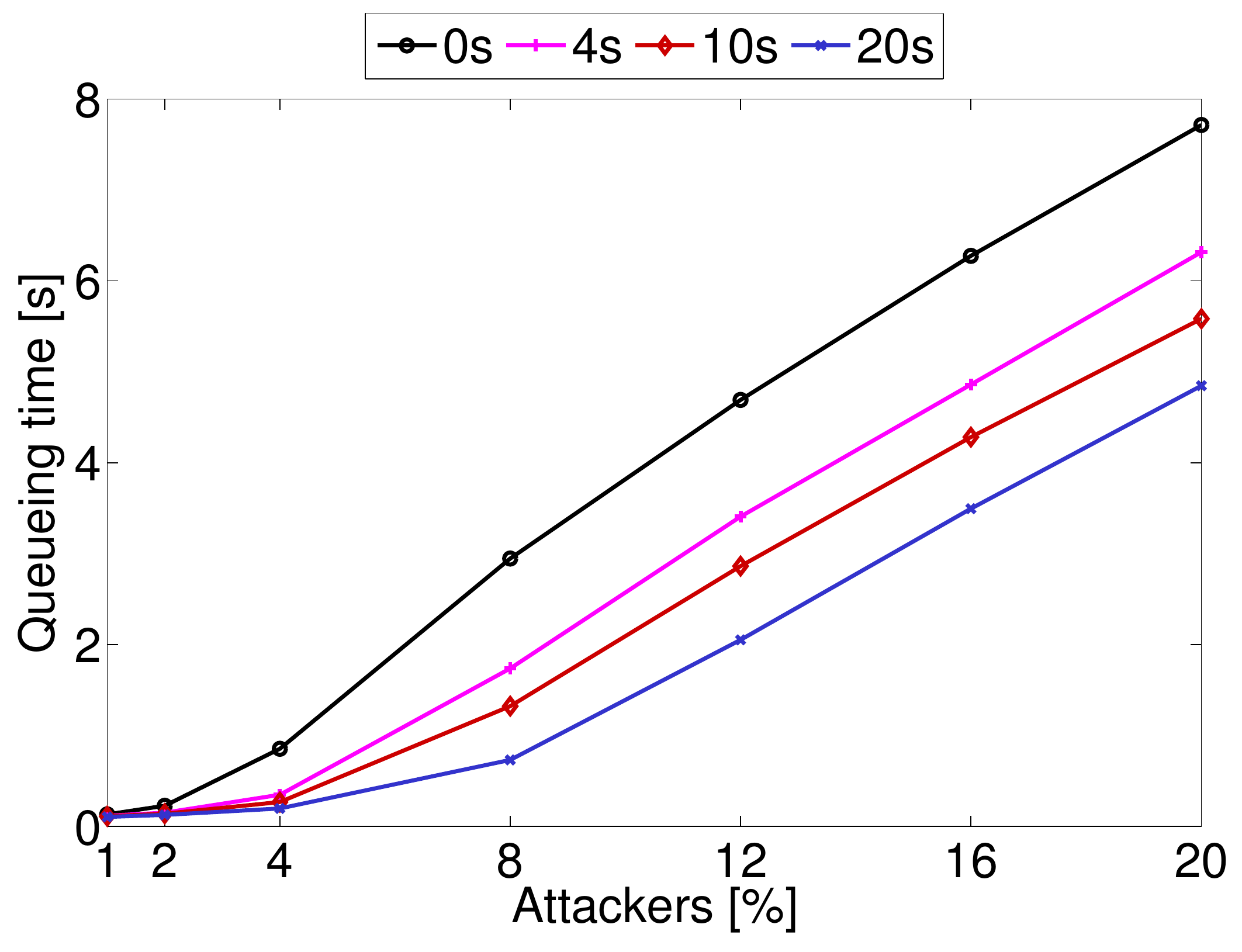}
		\label{fig:dchNoPchQTime}
	}
	\caption{Effect of DCH attacks on application response time and queueing time at the RNC signalling server, PCH disabled}
	\label{fig:dchRTQT}
\end{figure}

\begin{figure}[tbp]
	\centering
	\subfloat[Time spent in the FACH state while idle and busy] {
		\includegraphics[width=0.45\linewidth]{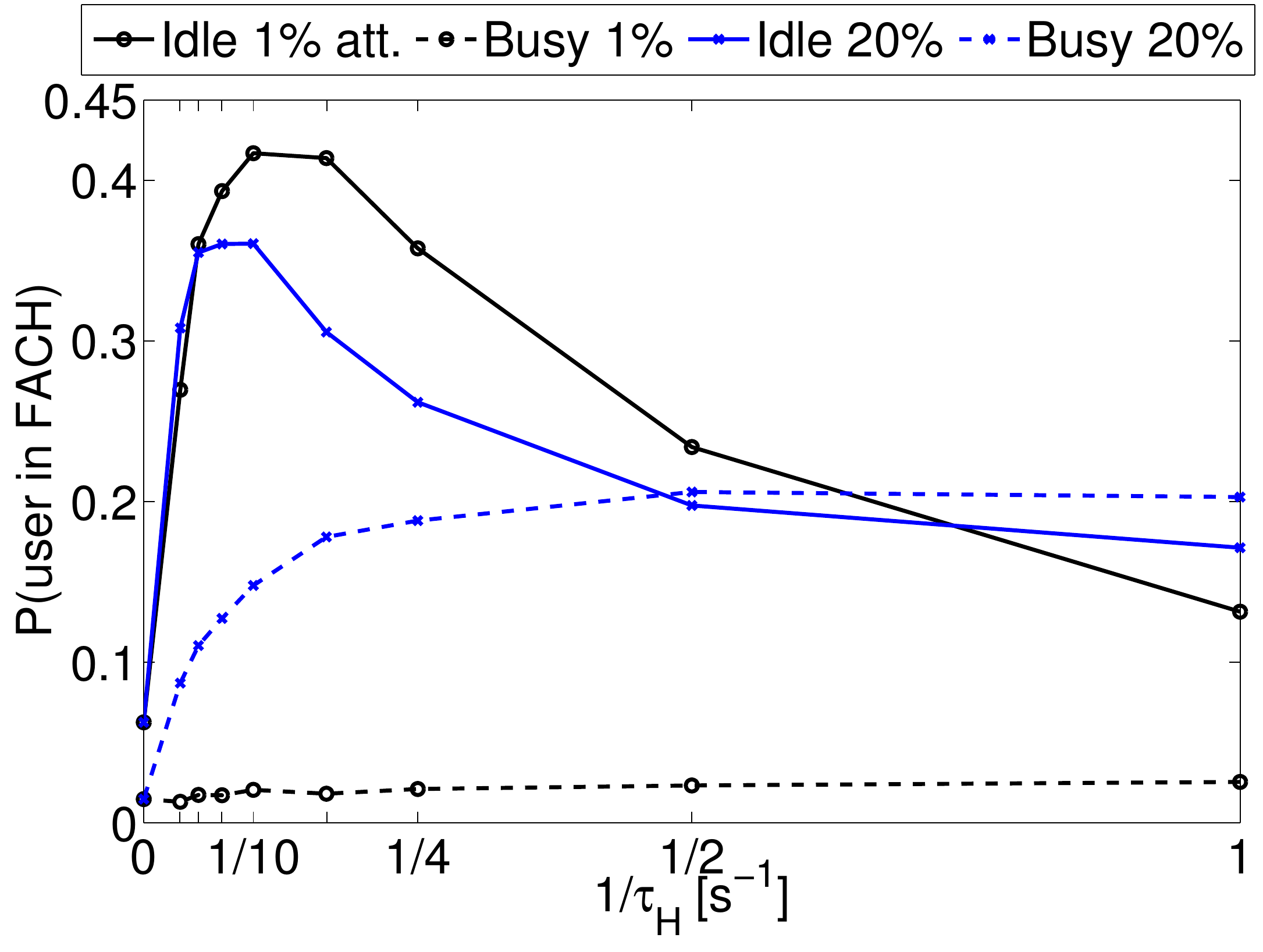}
		\label{fig:dchNoPchUtilFach}
	} \hspace{0.2cm}
	\subfloat[Time spent in the DCH state while idle and busy] {
		\includegraphics[width=0.45\linewidth]{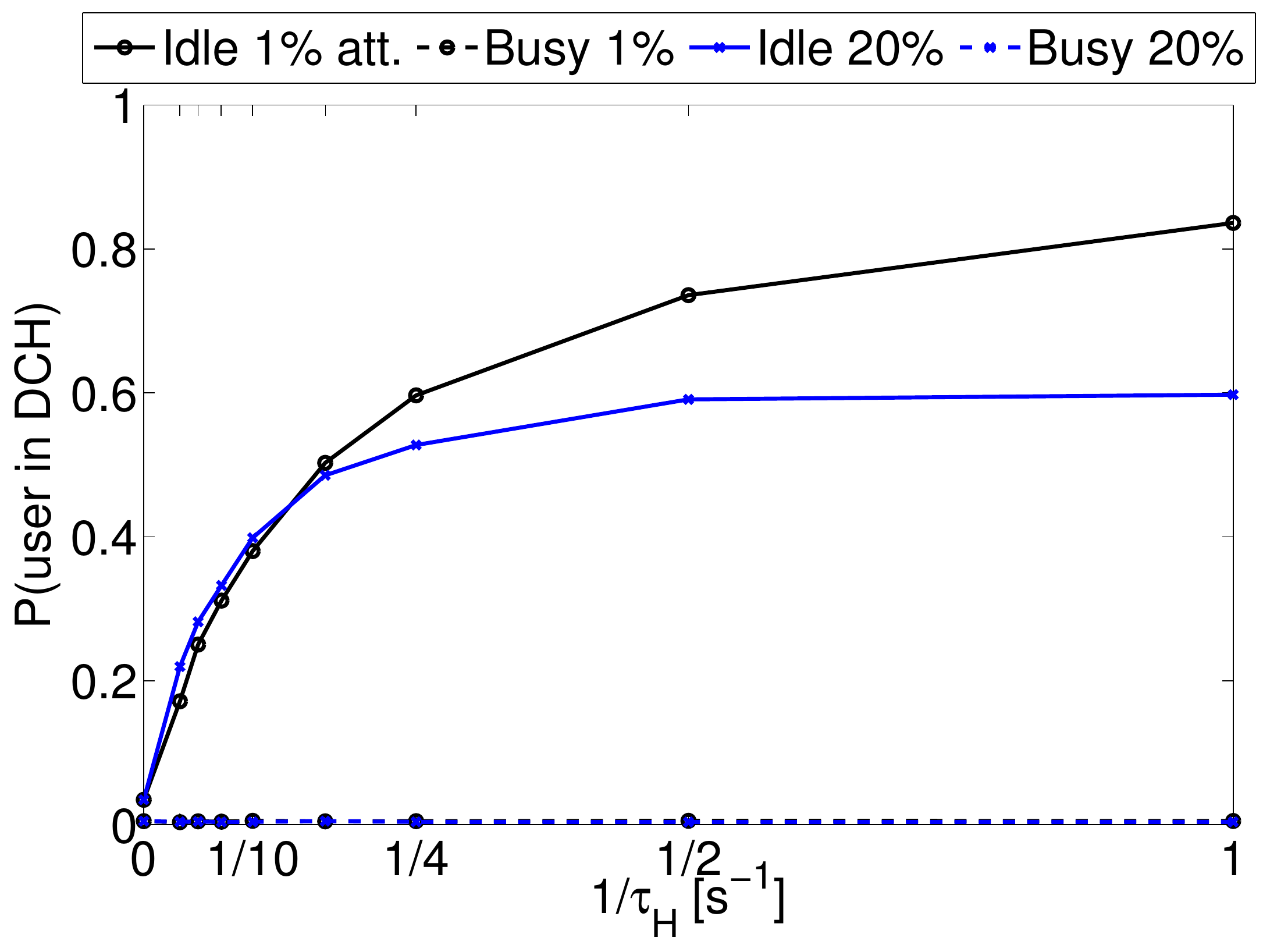}
		\label{fig:dchNoPchUtilDch}
	}
	\caption{Radio channel utilization under DCH attacks, PCH disabled}
	\label{fig:dchNoPchUtil}
\end{figure}

We performed simulation experiments in order to investigate the effect of signalling attacks and storms due to the RRC protocol on the RAN and the CN. We vary the number of compromised or misbehaving UEs from 1\% to 20\% of all UEs. Both normal and misbehaving UEs generate \textit{normal traffic} based on the web browsing model described above. The misbehaving applications are activated gradually between 20 and 30 minutes from the start of the simulation in order to prevent artifacts such as a huge spike of signalling load due to many malicious applications coming online at the same time. We collect simulation data only from the period when all misbehaving UEs are active. Each data point in the presented results is an average of five simulation runs with different random seeds. The relevant RRC protocol parameters are as given in Table~\ref{tab:rrcTransitions}. We also present analytical results derived from our mathematical model together with the simulation results. We observe that as a result of correctly adjusting the parameters of the mathematical model based on initial simulation results, and with the addition of the effect of congestion into the model, the simulation and analytical results show a high degree of agreement. We do not present analytical results for Figs.~\ref{fig:dchNoPchQTime} and \ref{fig:dchNoPchUtil} to prevent repetition of similar results, and for Fig.~\ref{fig:dchNoPchResponseTime} since the mathematical model does not capture quality-of-experience.

\begin{figure}[tbp]
	\centering
	\subfloat[PCH enabled (simulation)] {
		\includegraphics[width=0.45\linewidth]{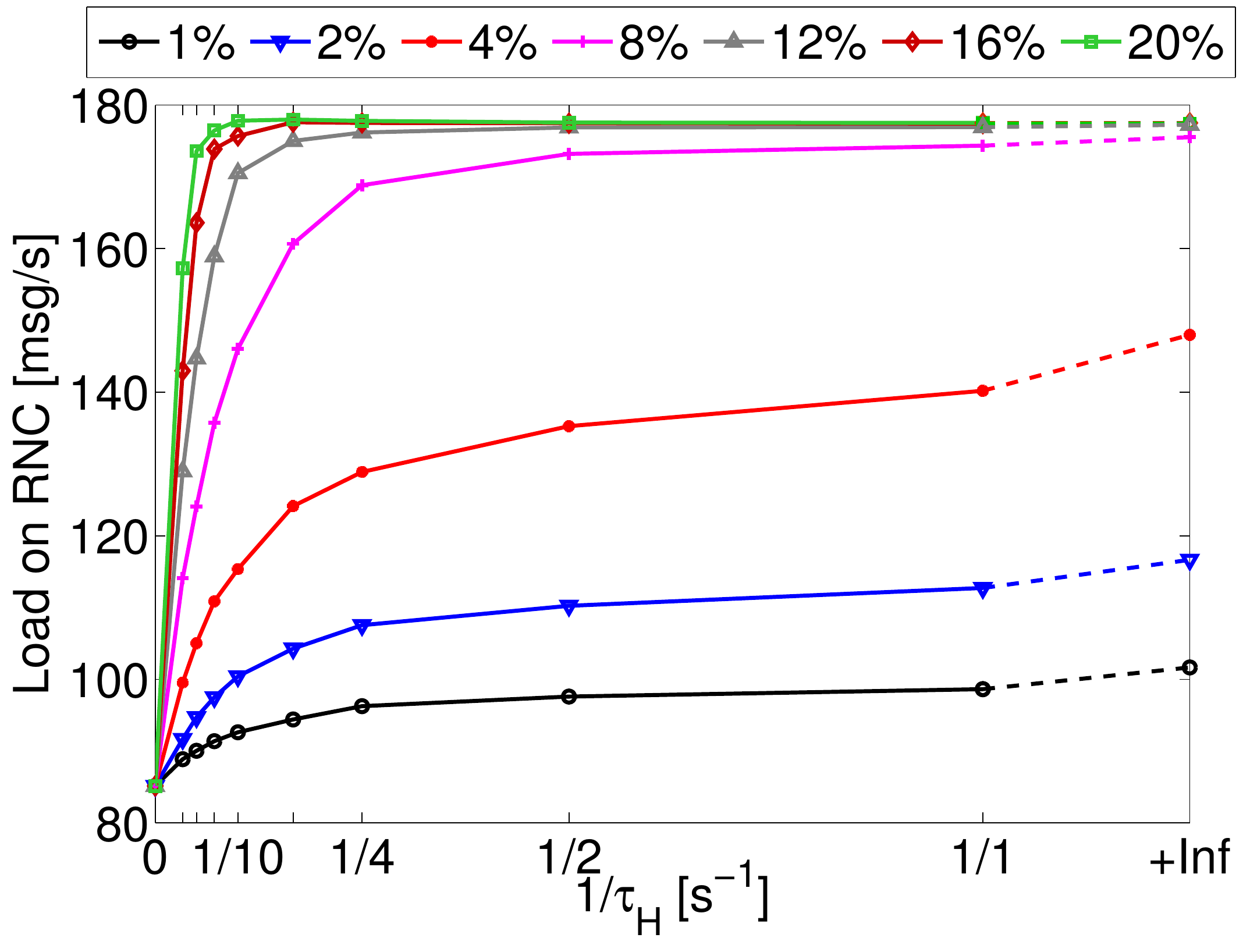}
		\label{fig:dchPchLoadRnc}
	} \hspace{0.2cm}
	\subfloat[PCH disabled (simulation)] {
		\includegraphics[width=0.45\linewidth]{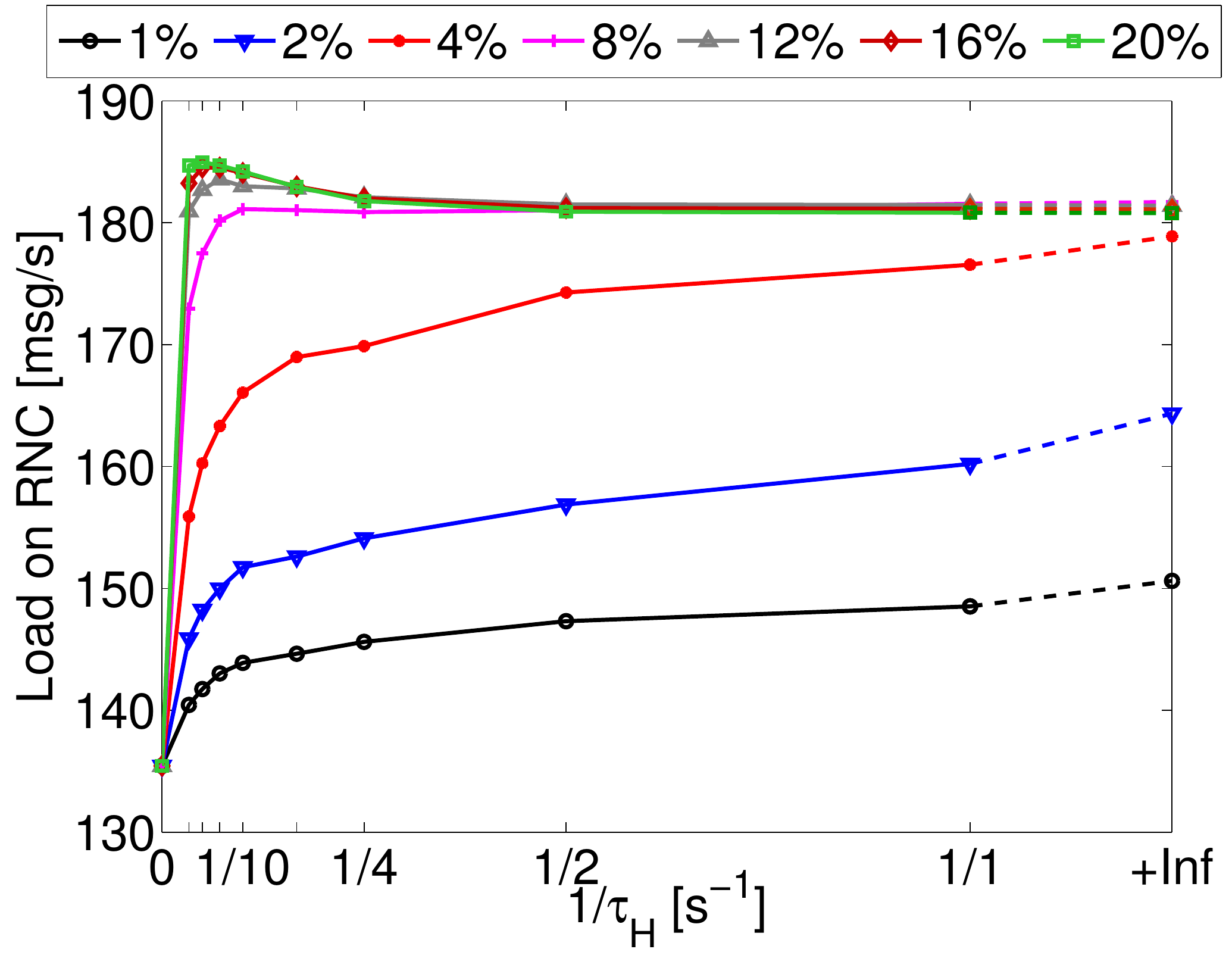}
		\label{fig:dchNoPchLoadRnc}
	}
	\\
	\subfloat[PCH enabled (analytical)] {
		\includegraphics[width=0.45\linewidth]{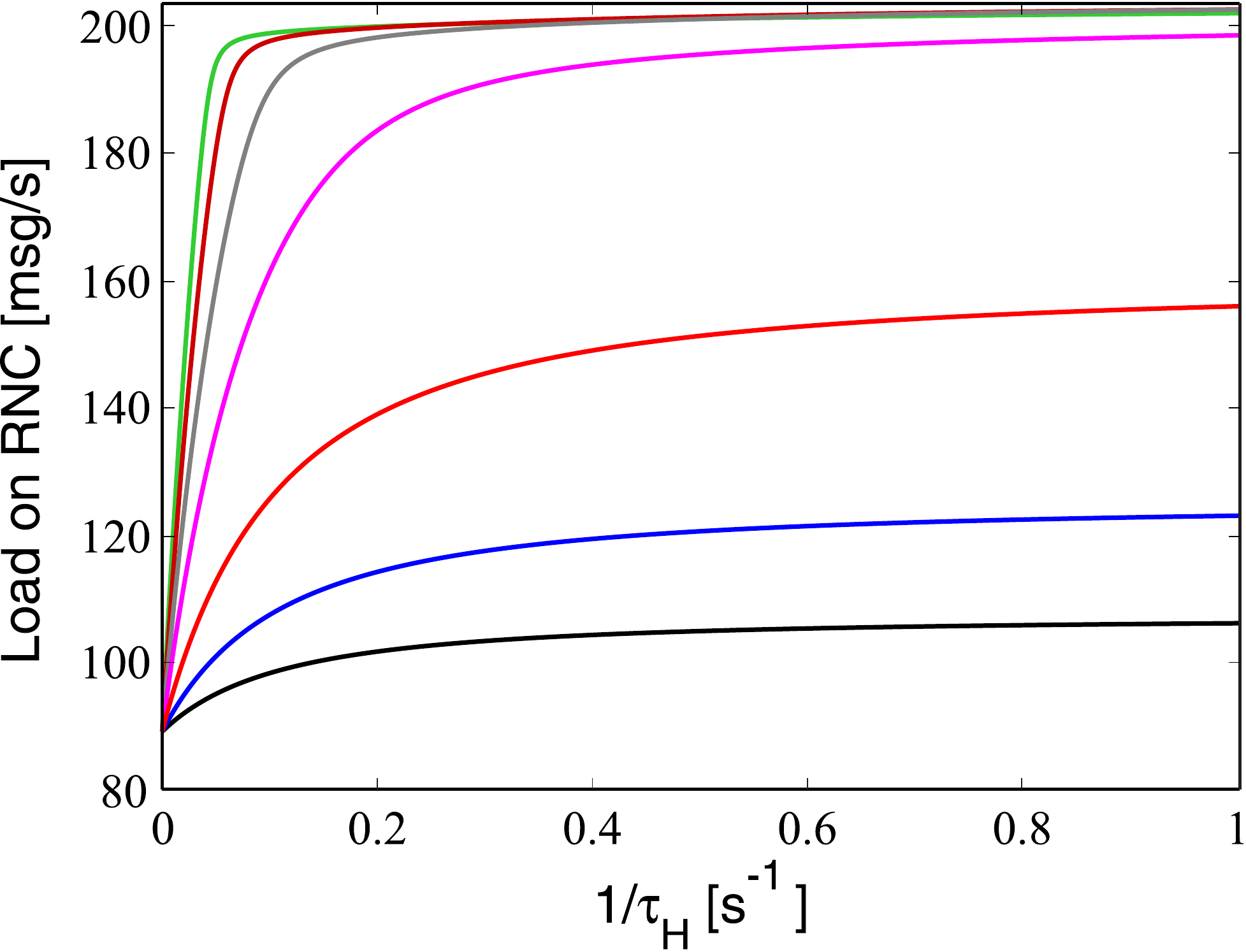}
		\label{fig:dchPchLoadRncMath}
	} \hspace{0.2cm}
	\subfloat[PCH disabled (analytical)] {
		\includegraphics[width=0.45\linewidth]{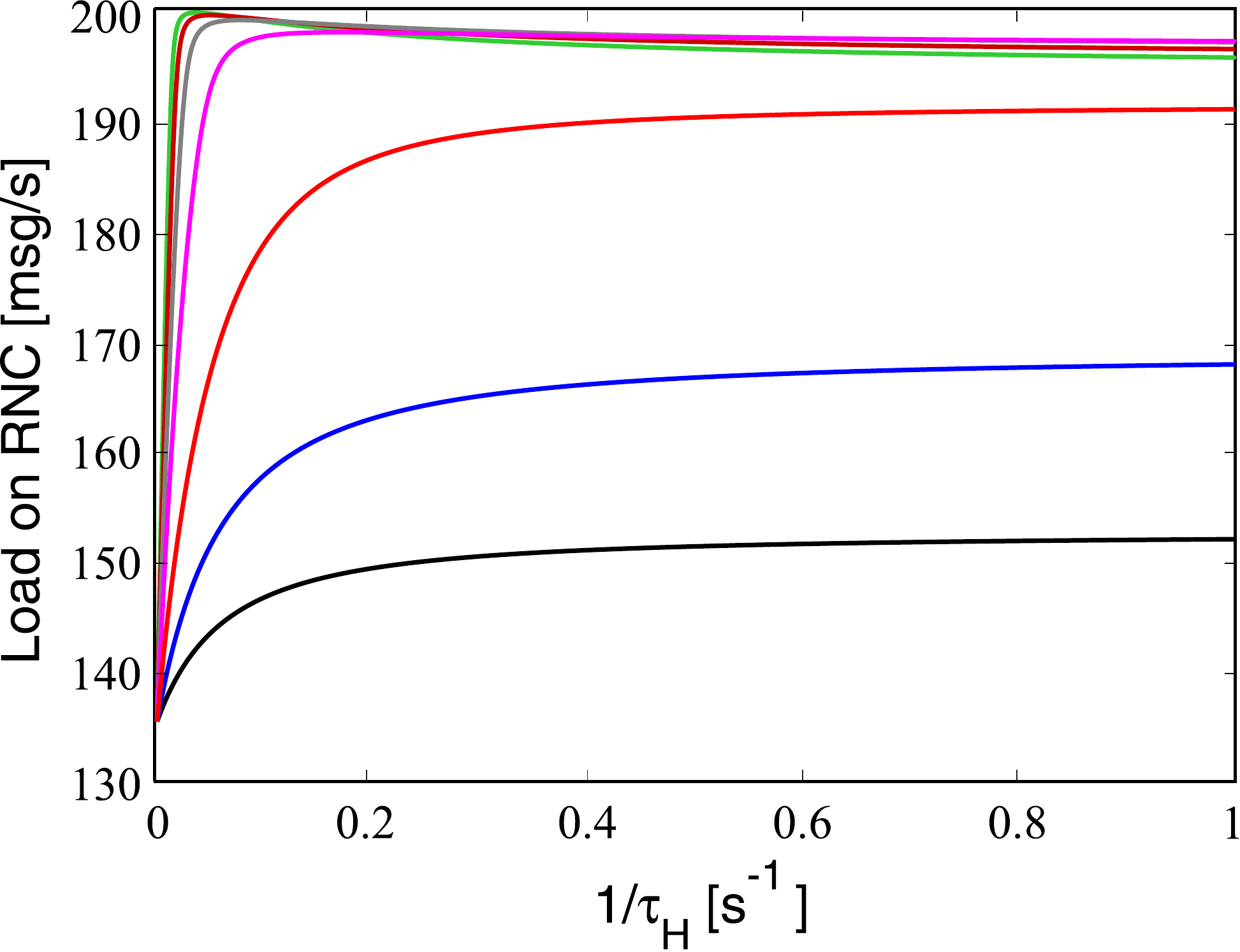}
		\label{fig:dchNoPchLoadRncMath}
	}
	\caption{Signalling load in the radio access network as a result of DCH attacks, for different number of attackers. The $1/\tau_H = 0$ case corresponds to a ``no attack'' scenario.}
	\label{fig:dchLoadRnc}
\end{figure}

Figure \ref{fig:dchLoadRnc} shows the signalling load in the RAN under DCH attacks, with PCH enabled and disabled. As $\tau_H$ decreases or the number of attackers increase, the number of signalling messages sent and received by the RNC towards the RAN increases as expected. The rate of increase is dependent on $1/\tau_H$ and higher when the number of attackers is high. We can see that whether the PCH state is enabled does not affect the behaviour of the signalling load in the RAN significantly, but it still decreases the signalling load. An interesting observation is that when PCH is disabled, there is a maximum load when the percentage of attackers is $\geq$ 8\% that is attained with a high $\tau_H$. This is worrying since it shows that a maximum signalling load can be induced in the RAN by signalling storms when a sufficient number of UEs misbehave without requiring a high level of synchronization between the misbehaving application and the RRC state machine. Enabling the PCH state addresses this issue. Another useful observation is that given a fixed number of attackers, RRC attacks are \textit{self-limiting}: as signalling load on the RNC increases, this prevents attackers from being able to attack the network at a high rate since they are themselves subject to longer waits for channel allocations. We will re-visit this issue when we discuss congestion at the RNC signalling server.

%----------------------------
%----------------------------

\begin{figure}[tbp]
	\centering
	\subfloat[PCH enabled (simulation)] {
		\includegraphics[width=0.45\linewidth]{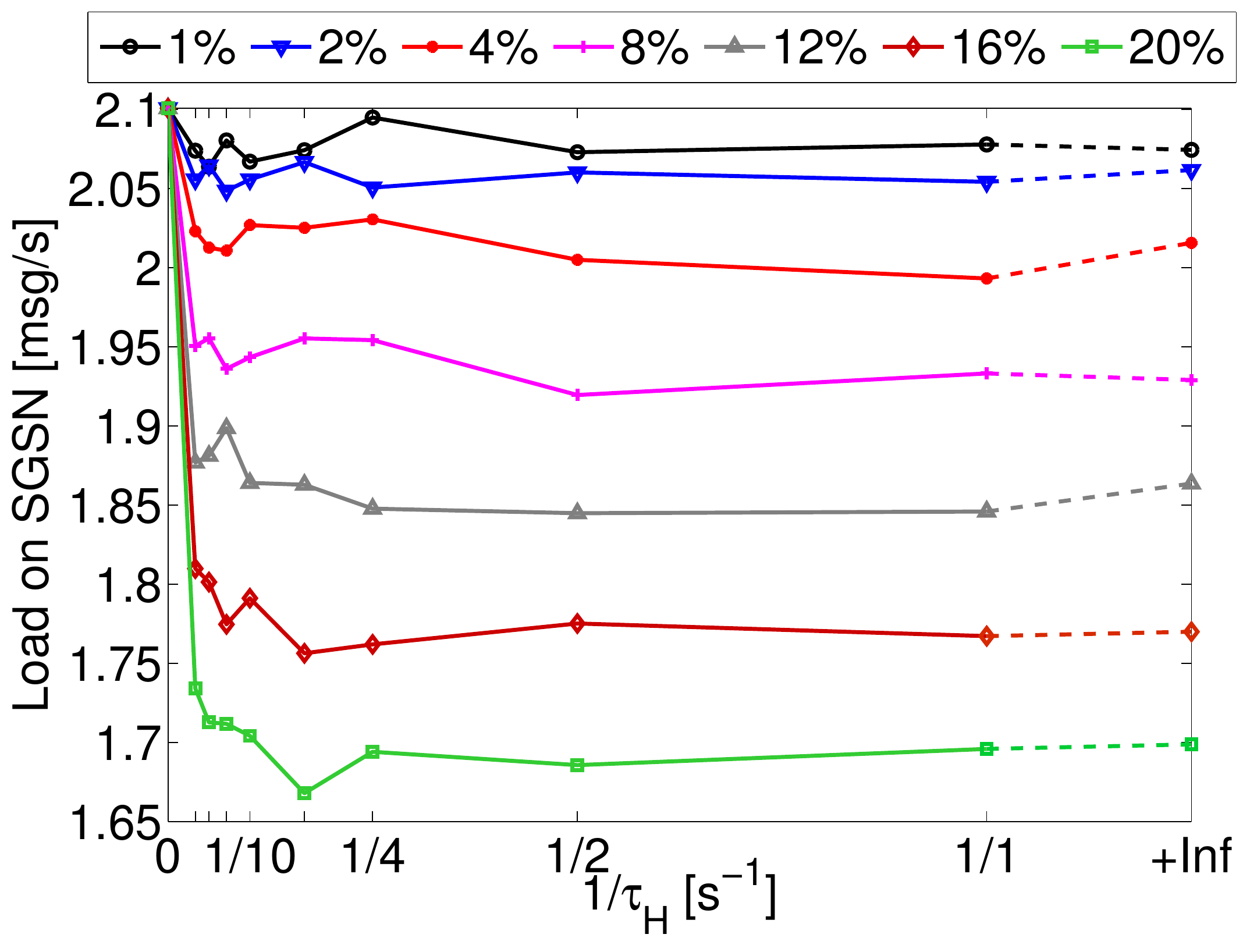}
		\label{fig:dchPchLoadSgsn}
	} \hspace{0.2cm}
	\subfloat[PCH disabled (simulation)] {
		\includegraphics[width=0.45\linewidth]{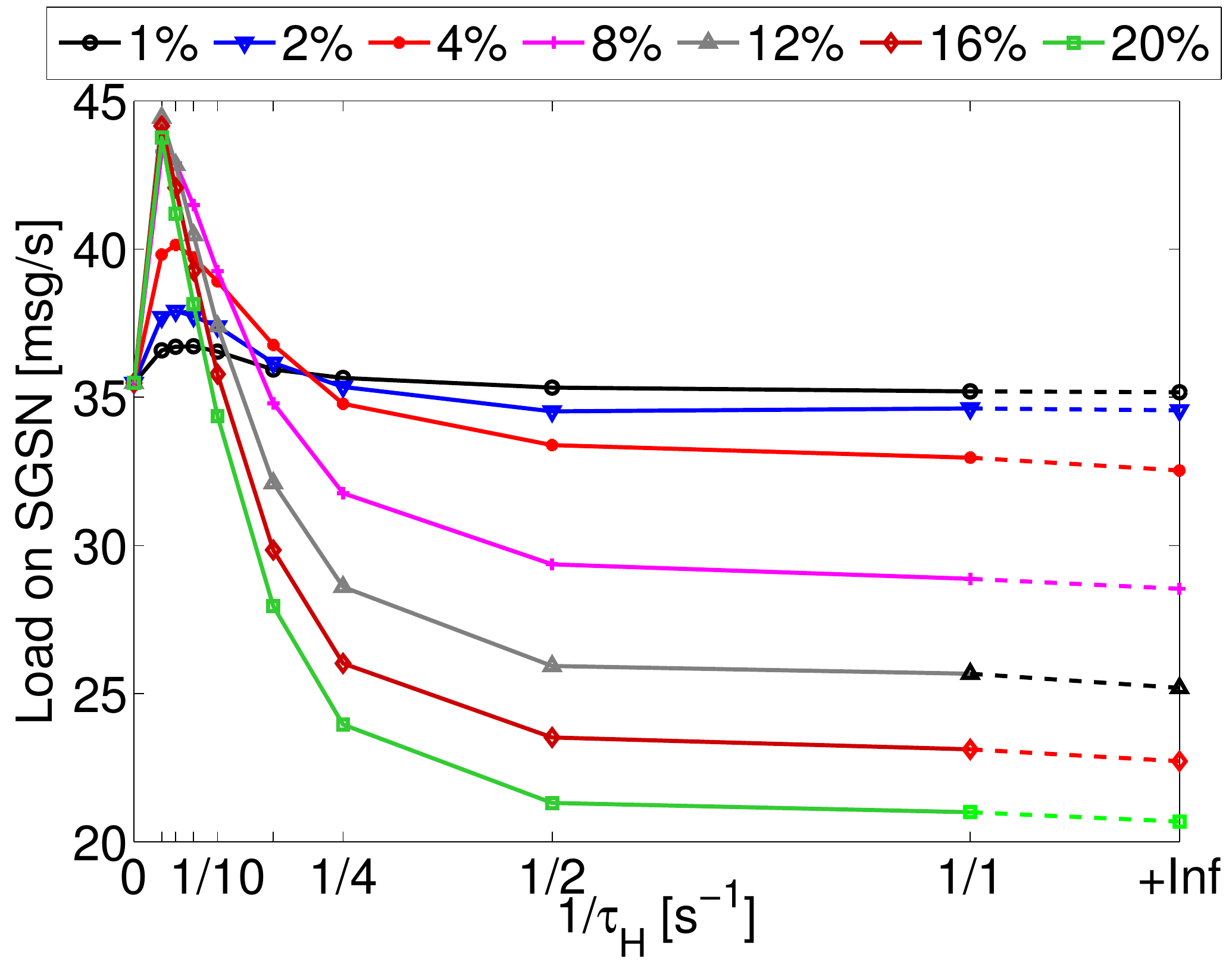}
		\label{fig:dchNoPchLoadSgsn}
	}
	\\
	\subfloat[PCH enabled (analytical)] {
		\includegraphics[width=0.45\linewidth]{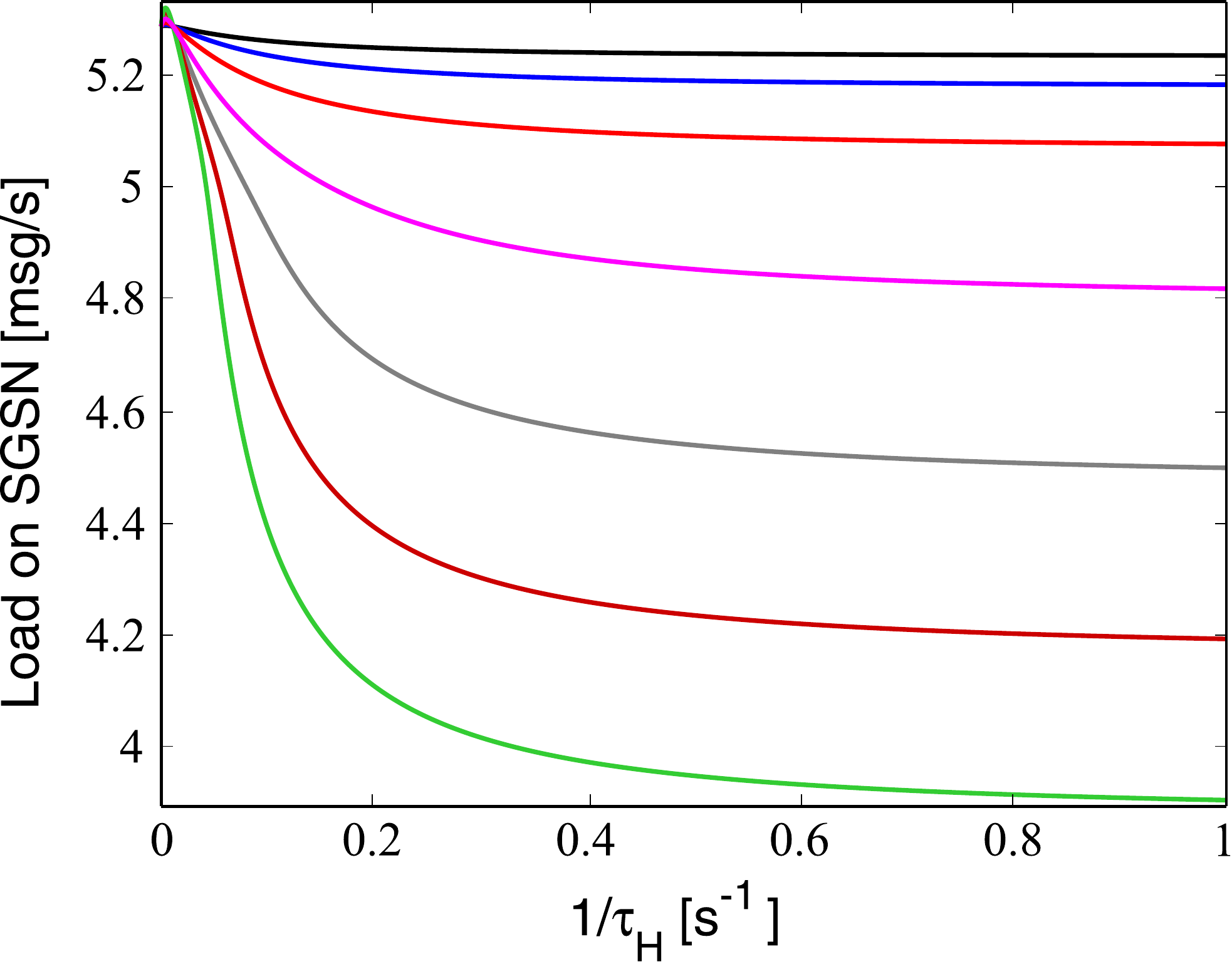}
		\label{fig:dchPchLoadSgsnMath}
	} \hspace{0.2cm}
	\subfloat[PCH disabled (analytical)] {
		\includegraphics[width=0.45\linewidth]{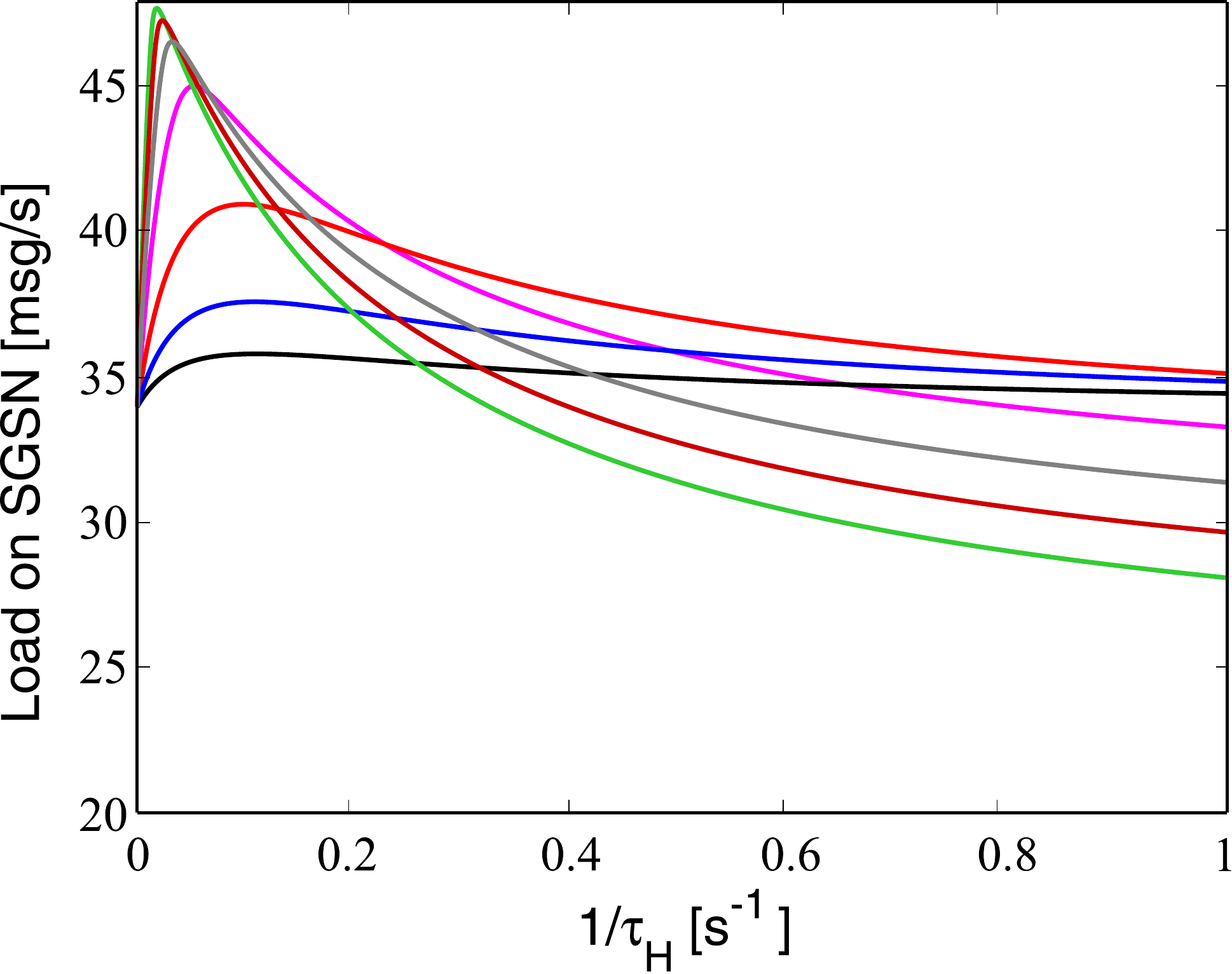}
		\label{fig:dchNoPchLoadSgsnMath}
	}
	\caption{Signalling load in the core network as a result of DCH attacks, for different number of attackers. The $1/\tau_H = 0$ case corresponds to a ``no attack'' scenario.}
	\label{fig:dchLoadSgsn}
\end{figure}

Figure \ref{fig:dchLoadSgsn} shows the signalling load in the CN under DCH attacks, with PCH enabled and disabled, and demonstrates the advantage of enabling the optional PCH state. We observe that whether the PCH state is enabled has a significant effect on the signalling load in the CN. This is because most RRC induced signalling with the CN occurs when the UE enters and exits the \textit{idle} state. Enabling the PCH state, which normally has a very long inactivity timer ($T_3$), prevents the UE from entering the idle state prematurely, significantly decreasing the signalling load in the CN at the cost of slightly more energy consumption at the UE. Therefore, our recommendation would be to enable PCH as a first step in the mitigation of RRC based signalling attacks and storms. Enabling the PCH state also eliminates the problem of the maximum signalling load observed in Fig.~\ref{fig:dchNoPchLoadSgsn} for high values of $\tau_H$, which is due to the interaction between $\tau_H$ and the RRC inactivity timers $T_1$ and $T_2$. When $\tau_H > T_1 + T_2$, the UE enters the \textit{idle} state as a result of inactivity, and then the misbehaving application causes the UE to go into FACH or DCH in order to send data, resulting in excessive signalling with the CN. The long $T_3$ timer of the PCH state solves this issue.

Our results so far demonstrate how the mobile network infrastructure is seriously affected by RRC based signalling anomalies. These anomalies also have an appreciable impact on the quality-of-experience (QoE) of the mobile user. Figure \ref{fig:dchNoPchResponseTime} shows the application response time at a normal UE, which is defined as the time between when the user requests a web page and when all of the web page is received. The response time is not greatly affected when there are very few misbehaving UEs and when $\tau_H$ is high. But delay increases by up to 400\% as the severity of the attack increases with increasing number of attackers and $1/\tau_H$. User normally tolerate a wait of 2--10 seconds for a web page to download~\cite{bib:nielsenUsability93,bib:nahWebWait04}, and therefore the observed response times are significant from a QoE view. The affected mobile users are highly likely to attribute the bad QoE to the MNO, so the MNO has one more incentive to detect and mitigate signalling problems in its network.

%----------------------------
%----------------------------

%----------------------------
%----------------------------

The main reason for the increase in application response time is the time it takes for the UE to acquire a radio channel in order to send and receive data, which includes, in addition to communication delays between the UE and the RNC, the service and queueing times experienced by the RRC signalling messages effecting the channel acquisition. Figure \ref{fig:dchNoPchQTime} shows that queueing time at the RRC signalling server component of the RNC greatly increases as the number of attackers increase. We observe that effects of congestion at the server become significant when the percentage of attackers is $>$ 4\%, affecting application response time for normal users, and also placing a limit on the impact of signalling attacks on the network since the attackers themselves are subject to longer delays for channel acquisition.

%----------------------------
%----------------------------

Our final results relate to how the UE utilizes its allocated radio resources, and provide a useful feature that we aim to exploit in our future work on the detection of signalling attacks. Figure \ref{fig:dchNoPchUtil} shows the ratio of time the UE is in the FACH or DCH state while busy (i.e. sending or receiving data) and idle. The most important observation is that a normal UE, represented with $1/\tau_H = 0$, has a markedly different behaviour than a misbehaving UE ($1/\tau_H > 0$), and the discrepancy increases with $1/\tau_H$. Normal UEs do not spend a significant time in FACH or DCH as busy or idle, but attackers spend a long time as idle while in FACH and DCH, i.e. their session tails are comparatively \textit{longer} than their session body. This is because normal users only acquire the channel when they have legitimate traffic, and they send larger chunks of data and therefore use the channel for longer than attackers, resulting in a low ratio of idle to busy time. Attackers, on the other hand, frequently acquire the channel to send only a small amount of ``attack traffic'' and therefore waste most of the radio channel as reflected in their high ratio of idle to busy time. The exception to this is the FACH state when there is congestion in the control plane due to the signalling attack: we observe that attackers spend significantly long times as busy in the FACH state when there is congestion, e.g. with 20\% of attackers, which is due to the long delay it takes the UEs to acquire the channel as discussed above.

\section{Conclusions and Future Work}
\label{sec:conclusion}

In this paper, we have investigated the effect of signalling attacks and storms in mobile networks, focusing on signalling anomalies that exploit the radio resource control (RRC) protocol. We presented a Markov model of the signalling behaviour of the UE and extended the model for effects of congestion in the control plane. The analytical model provides an accurate representation of the RRC signalling behaviour and allows us to reach quick analytical results.

We have also developed a simulation of a UMTS mobile network, and simulation experiments were used to validate the mathematical model, resulting in its improvement by the addition of concepts not previously captured and the realistic setting of the model parameters. We presented simulation and analytical results, looking at how different components in the mobile network are affected by signalling attacks and storms.

Our results show that RRC based signalling anomalies can cause significant problems in both the control plane and the user plane in the network, and provide insight into how such attacks and storms can be detected and mitigated. While we have focused on UMTS networks in this work, the RRC protocol is also employed in LTE networks, and any RRC related anomalies would have a more severe impact in LTE networks since they employ only two RRC states (connected and idle), and the mitigating effect of the long $T_3$ timer used in the PCH state are non-existent in LTE networks.

Future work can exploit the insight gained in this paper for the detection and mitigation of signalling attacks in mobile networks. One aspect that requires attention is the identification of possible locations, such as specific cells, where attacks may originate, and methods related to search and smart traffic routing may prove valuable in this context \cite{Cao,Zarina}.  Another important aspect relates to identifying sets of representative features for the detection of signalling attacks and storms, and of the misbehaving UEs. An important consideration is to prevent false positives as much as possible so as not to punish normal ``heavy'' users. We will also develop system wide models based on queueing theory \cite{G-nets} that represent a single user in a simple manner, to study mitigation methods that involve randomisation and adaptively introducing artificial delays in the state transitions of the UEs so that they may automatically reduce the negative impact of attacks and signalling storms.

\section*{Acknowledgments}
 
The work presented in this paper was partially supported by the EU FP7 research project NEMESYS (Enhanced Network Security for Seamless Service Provisioning in the Smart Mobile Ecosystem), under grant agreement no. 317888 within the FP7-ICT-2011.1.4 Trustworthy ICT domain.

%----------------------------------------

\IEEEtriggeratref{38}

\bibliographystyle{IEEEtran}
\bibliography{references}


% \begin{IEEEbiography}{Gokce Gorbil}
% \end{IEEEbiography}
% 
% \begin{IEEEbiography}{Omer H. Abdelrahman}
% \end{IEEEbiography}
% 
% \begin{IEEEbiography}{Mihajlo Pavloski}
% \end{IEEEbiography}
% 
% \begin{IEEEbiography}{Erol Gelenbe}
% \end{IEEEbiography}


%\newpage


%\vfill

% is flush with the other column:

\end{document}